\documentclass[final,5p,times,twocolumn]{elsarticle}

\usepackage{graphics}
\graphicspath{{fig/}}
\DeclareGraphicsExtensions{.pdf,.jpeg,.png}
\usepackage{mdwmath}
\usepackage{mdwtab}
\usepackage{rotating}
\usepackage[cmex10]{amsmath}
\usepackage{stfloats}
\usepackage[font=small]{caption}
\usepackage[justification=centering]{caption}
\usepackage{multirow}
\usepackage{amssymb}
\usepackage{xcolor}
\emergencystretch=\maxdimen

 \biboptions{numbers,sort&compress}

\begin{document}

\begin{frontmatter}



\title{An Empirical Study on Software Defect Prediction with a Simplified Metric Set}


\author[a,b]{Peng He}
\author[c,d]{Bing Li}
\author[a,b]{Xiao Liu}
\author[b,e]{Jun Chen}
\author[b,d]{Yutao Ma \corref{cor1}}

\address[a]{State Key Laboratory of Software Engineering, Wuhan University, Wuhan 430072, China}
\address[b]{School of Computer, Wuhan University, Wuhan 430072, China}
\address[c]{International School of Software, Wuhan University, Wuhan 430079, China}
\address[d]{Research Center for Complex Network, Wuhan University, Wuhan 430072, China}
\address[e]{National Engineering Research Center for Multimedia Software, Wuhan University, Wuhan 430072, China}
\cortext[cor1]{Corresponding author. Tel: +86 27 68776081 \\   E-mail: \{penghe (P. He), bingli (B. Li), lxiao (X. Liu), chenj (J. Chen), ytma (Y.T. Ma)\}@whu.edu.cn}

\begin{abstract}
\emph{Context:} Software defect prediction plays a crucial role in estimating the most defect-prone components of software, and a large number of studies have pursued improving prediction accuracy within a project or across projects. However, the rules for making an appropriate decision between within- and cross-project defect prediction when available historical data are insufficient remain unclear.  \\
\emph{Objective:} The objective of this work is to validate the feasibility of the predictor built with a simplified metric set for software defect prediction in different scenarios, and to investigate practical guidelines for the choice of training data, classifier and metric subset of a given project. \\
\emph{Method:} First, based on six typical classifiers, three types of predictors using the size of software metric set were constructed in three scenarios. Then, we validated the acceptable performance of the predictor based on Top-\emph{k} metrics in terms of statistical methods. Finally, we attempted to minimize the Top-\emph{k} metric subset by removing redundant metrics, and we tested the stability of such a minimum metric subset with one-way ANOVA tests. \\
\emph{Results:} The study has been conducted on 34 releases of 10 open-source projects available at the PROMISE repository. The findings indicate that the predictors built with either Top-\emph{k} metrics or the minimum metric subset can provide an acceptable result compared with benchmark predictors. The guideline for choosing a suitable simplified metric set in different scenarios is presented in Table \ref{summary2}. \\
\emph{Conclusion:} The experimental results indicate that (1) the choice of training data for defect prediction should depend on the specific requirement of accuracy; (2) the predictor built with a simplified metric set works well and is very useful in case limited resources are supplied; (3) simple classifiers (e.g., Na\"{\i}ve Bayes) also tend to perform well when using a simplified metric set for defect prediction; (4) in several cases, the minimum metric subset can be identified to facilitate the procedure of general defect prediction with acceptable loss of prediction precision in practice.

\end{abstract}


\end{frontmatter}


\section{Introduction}
In software engineering, defect prediction can precisely estimate the most defect-prone software components, and help software engineers allocate limited resources to those bits of the systems that are most likely to contain defects in testing and maintenance phases. Understanding and building defect predictors (also known as defect prediction models) for a software project is useful for a variety of software development or maintenance activities, such as assessing software quality and monitoring quality assurance (QA).

The importance of defect prediction has motivated numerous researchers to define different types of models or predictors that characterize various aspects of software quality. Most studies usually formulate such a problem as a supervised learning problem, and the outcomes of those defect prediction models depend on historical data. That is, they trained predictors from the data of historical releases in the same project and predicted defects in the upcoming releases, or reported the results of cross-validation on the same data set \cite{He:An}, which is referred to as Within-Project Defect Prediction (WPDP). Zimmermann \emph{et al.} \cite{Zimmermann:Cross} stated that defect prediction performs well within projects as long as there is a sufficient amount of data available to train any models. However, it is not practical for new projects to collect such sufficient historical data. Thus, achieving high accuracy defect prediction based on within-project data is impossible in some cases.

Conversely, there are many public on-line defect data sets available, such as PROMISE\footnote{http://promisedata.org}, Apache\footnote{http://www.apache.org/} and Eclipse\footnote{http://eclipse.org}. Some researchers have been inspired to overcome this challenge by applying the predictors built for one project to a different one  \cite{Zimmermann:Cross,Briand:Assessing,Ma: Transfer}. Utilizing data across projects to build defect prediction models is commonly referred to as Cross-Project Defect Prediction (CPDP). CPDP refers to predicting defects in a project using prediction models trained from the historical data of other projects. The selection of training data depends on the distributional characteristics of data sets. Some empirical studies evaluated the potential usefulness of cross-project predictors with a number of software metrics (e.g., static code metrics, process metrics, and network metrics) \cite{Rahman:Recalling,He:An} and how these metrics could be used in a complementary manner \cite{Premraj:Network}. Unfortunately, despite these attempts to demonstrate the feasibility of CPDP, this method has been widely challenged because of its low performance in practice \cite{Rahman:Recalling}. Moreover, it is still unclear how defect prediction models between WPDP and CPDP are rationally chosen when limited or insufficient historical data are provided.

\begin{figure*}
  \centering
  \includegraphics[width=6in]{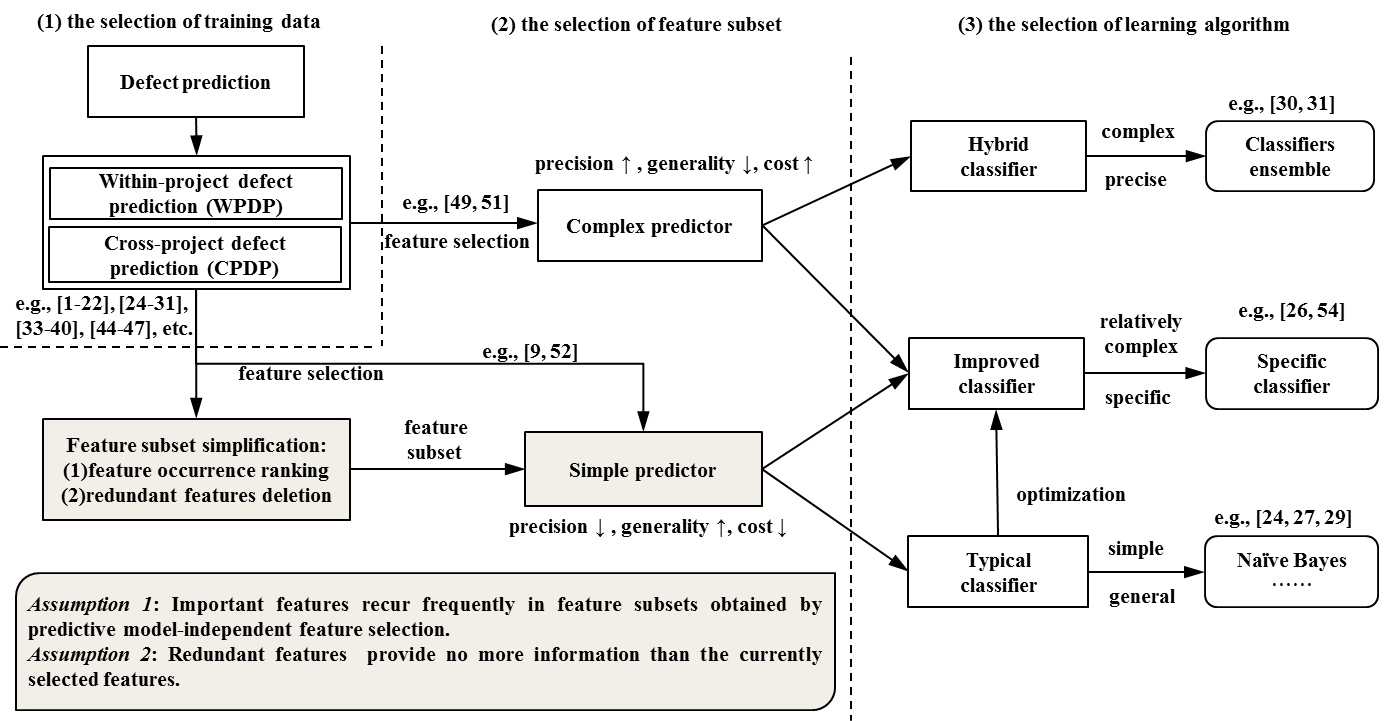}\\
  \caption{A summary of the state-of-the-art defect prediction.}\label{Fig.architectre}
\end{figure*}

 In defect prediction literature, a considerable number of software metrics, such as static code metrics, code change history, process metrics and network metrics \cite{He:Using}, have been used to construct different predictors for defect prediction \cite{Radjenovic:Software}. Almost all existing prediction models are built on the complex combinations of software metrics, with which a prediction model usually can achieve a satisfactory accuracy. Although some feature selection techniques (e.g., principal component analysis (PCA)) successfully reduce data dimensions \cite{Zimmermann:Predicting,Turhan: Analysis,Shivaji: Reducing,Lu:Software,Wang:A}, they still lead to a time-consuming prediction process. Can we find a compromise solution that makes a tradeoff between cost and accuracy? In other words, can we find a universal predictor built with few metrics (e.g., Lines of Code (LOC)) that achieves an acceptable result compared with those complex prediction models?

In addition to the selection of a wide variety of software metrics, there are many classifiers (learning algorithms) that have been studied, such as  Na\"{\i}ve Bayes, J48, Support Vector Machine (SVM), Logistic Regression, and Random Tree \cite{Jin:Applications,Catal: Investigating,Song:Software}, and defect prediction using these typical classifiers has achieved many useful conclusions. Currently, some improved classifiers \cite{Arisholma:A,Jiang:Personzlied,Ma: Transfer} and hybrid classifiers \cite{WANG:Software,Tosun:Ensemble} have also been proposed to effectively improve classification results. Menzies \emph{et al.} \cite{Menzies:Defect} advocated that different classifiers have indiscriminate usage and must be chosen and customized for the goal at hand.

Figure \ref{Fig.architectre} presents a summary of the state-of-the-art defect prediction. Complex predictors improve prediction precision with loss of generality and increase the cost of data acquisition and processing. On the contrary, simple predictors are more universal, and they reduce the total effort-and-cost by sacrificing a little precision. To construct an appropriate and practical prediction model, we should take into overall consideration the precision, generality and cost according to specific requirements. Unlike the existing studies on complex predictors, in our study, we focus mainly on building simple prediction models with a simplified metric set according to two assumptions (see the contents with a gray background in Figure \ref{Fig.architectre}), and seek empirical evidence that they can achieve acceptable results compared with the benchmark models. Our contributions to the current state of research are summarized as follows:
\begin{itemize}
\item We proposed an easy-to-use approach to simplifying the set of software metrics based on \emph{filters} methods for feature selection, which could help software engineers build suitable prediction models with the most representative code features according to their specific requirements.
\item We also validated the optimistic performance of the prediction model built with a simplified subset of metrics in different scenarios, and found that it was competent enough when using different classifiers and training data sets from an overall perspective.
\item We further demonstrated that the prediction model constructed with the minimum subset of metrics can achieve a respectable overall result. Interestingly, such a minimum metric subset is stable and independent of the classifiers under discussion.
\end{itemize}

With these contributions, we complement previous work on defect prediction. In particular, we provide a more comprehensive suggestion on the selection of appropriate predictive modeling approaches, training data, and simplified metric sets for constructing a defect predictor according to different specific requirements.

The rest of this paper is organized as follows. Section 2 is a review of related literature. Sections 3 and 4 describe the approach of our empirical study and the detailed experimental setups, respectively. Sections 5 and 6 analyze and discuss the primary results, and some threats to validity that could affect our study are presented in Section 7. Finally, Section 8 concludes the paper and presents the agenda for future work.

\section{Related Work}
Defect prediction is an important topic in software engineering, which allows software engineers to pay more attention to defect-prone code with software metrics, thereby improving software quality and making better use of limited resources.

\subsection{Within-Project Defect Prediction}
Catal \cite{Catal:Software} investigated 90 software defect prediction papers published between 1990 and 2009. He categorized these papers and reviewed each paper from the perspectives of metrics, learning algorithms, data sets, performance evaluation metrics, and experimental results in an easy and effective manner. According to this survey, the author stated that most of the studies using method-level metrics and prediction models were mostly based on machine learning techniques, and Na\"{\i}ve Bayes was validated as a robust machine learning algorithm for supervised software defect prediction problems.

Hall \emph{et al.} \cite{Hall:A} investigated how the context of models, the independent variables used, and the modeling techniques applied affected the performance of defect prediction models according to 208 defect prediction studies. Their results showed that simple modeling techniques, such as Na\"{\i}ve Bayes or Logistic Regression, tended to perform well. In addition, the combinations of independent variables were used by those prediction models that performed well, and the results were particularly good when feature selection had been applied to these combinations. The authors argued that there were a lot of defect prediction studies in which confidence was possible, but more studies that used a reliable methodology and that reported their detailed context, methodology, and performance in the round were needed.

The vast majority of these studies were investigated in the above two systematic literature reviews that were conducted in the context of WPDP. However, they ignored the fact that some projects, especially new projects, usually have limited or insufficient historical data to train an appropriate model for defect prediction. Hence, some researchers have begun to divert their attention toward CPDP.

\subsection{Cross-Project Defect Prediction}
To the best of our knowledge, the earliest study on CPDP was performed by Briand \emph{et al.} \cite{Briand:Assessing}, who applied models built on an open-source project (i.e., Xpose) to another one (i.e., Jwriter). Although the predicted defect detection probabilities were not realistic, the fault-prone class ranking was accurate. They also validated that such a model performed better than the random model and outperformed it in terms of class size. Zimmermann \emph{et al.} \cite{Zimmermann:Cross} conducted a large-scale experiment on data vs. domain vs. process, and found that CPDP was not always successful (21/622 predictions). They also found that CPDP was not symmetrical between Firefox and IE.

Turhan \emph{et al.} \cite{Turhan:On} analyzed CPDP using static code features based on 10 projects also collected from the PROMISE repository. They proposed a nearest-neighbor filtering technique to filter out the irrelevancies in cross-project data. Moreover, they further investigated the case where models were constructed from a mix of within- and cross-project data, and checked for any improvements to WPDP after adding the data from other projects. They concluded that when there was limited project historical data (e.g., $10\%$ of historical data), mixed project predictions were viable, as they performed as well as within-project prediction models \cite{Turhan:Empirical}.

Rahman \emph{et al.} \cite{Rahman:Recalling} conducted a cost-sensitive analysis of the efficacy of CPDP on 38 releases of nine large Apache Software Foundation (ASF) projects, by comparing it with WPDP. Their findings revealed that the cost-sensitive cross-project prediction performance was not worse than the within-project prediction performance, and was substantially better than random prediction performance. Peters \emph{et al.} \cite{Peters:Better} introduced a new filter to aid cross-company learning compared with the state-of-the-art Burak filter. The results revealed that their approach could build $64\%$ more useful predictors than both within-company and cross-company approaches based on Burak filters, and demonstrated that cross-company defect prediction was able to be applied very early in a project's lifecycle.

He \emph{et al.} \cite{He:An} conducted three experiments on the same data sets used in this study to validate the idea that training data from other projects can provide acceptable results. They further proposed an approach to automatically selecting suitable training data for projects without local data. Towards training data selection for CPDP, Herbold \cite{Herbold:Training} proposed several strategies based on 44 data sets from 14 open-source projects. Parts of their data sets are used in our paper. The results demonstrated that their selection strategies improved the achieved success rate significantly, whereas the quality of the results was still unable to compete with WPDP.

The review reveals that prior studies have mainly investigated the feasibility of CPDP and the choice of training data from other projects. However, relatively little attention has been paid to empirically exploring the performance of a predictor based on a simplified metric set from the perspectives of effort-and-cost, accuracy and generality. Moreover, very little is known about whether the predictors built with simplified or minimum software metric subsets obtained by eliminating some redundant and irrelevant features are able to achieve acceptable results.

\subsection{Software Metrics}
A wide variety of software metrics treated as features have been used for defect prediction to improve software quality. At the same time, numerous comparisons among different software metrics have also been made to examine which metric or combination of metrics performs better.

Shin \emph{et al.} \cite{Shin: Evaluating} investigated whether source code and development histories were discriminative and predictive of vulnerable code locations among complexity, code churn, and developer activity metrics. They found that 24 of the 28 metrics were discriminative for both Mozilla Firefox and Linux kernel. The models using all the three types of metrics together predicted over $80\%$ of the known vulnerable files with less than $25\%$ false positives for both projects. Marco \emph{et al.} \cite{Marco:Evaluating} conducted three experiments on five systems with process metrics, previous defects, source code metrics, entropy of changes, churn, etc. They found that simple process metrics were the best overall performers, slightly ahead of the churn of source code and the entropy of source code metrics.

Zimmermann and Nagappan \cite{Zimmermann:Predicting} leveraged social network metrics derived from dependency relationships between software entities on Windows Server 2003 to predict which entities were likely to have defects. The results indicated that network metrics performed significantly better than source code metrics with regard to predicting defects. Tosun \emph{et al.} \cite{Tosun:Validation} conducted additional experiments on five public data sets to reproduce and validate their results from two different levels of granularity. The results indicated that network metrics were suitable for predicting defects for large and complex systems, whereas they performed poorly on small-scale projects. To further validate the generality of the findings, Premraj and Herzig \cite{Premraj:Network} replicated Zimmermann and Nagappan's work on three open-source projects, and found that the results were consistent with the original study. However, with respect to the collection of data sets, code metrics might be preferable for empirical studies on open-source software projects.

Recently, Radjenovi\'{c} \emph{et al.} \cite{Radjenovic:Software} classified 106 papers on defect prediction according to metrics and context properties. They found that the proportions of object-oriented metrics, traditional source code metrics, and process metrics were $49\%$, $27\%$, and $24\%$, respectively. Chidamber and Kemerer's (CK) suite metrics are most frequently used. Object-oriented and process metrics have been reported to be more successful than traditional size and complexity metrics. Process metrics appear to be better at predicting post-release defects than any static code metrics. For more studies, one can refer to the literature \cite{Arisholma:A,Nagappan:Using,Menzies:Data,Yu:Experience}.

The simplification of software metric set could improve the performance and efficiency of defect prediction. Feature selection techniques have been used to remove redundant or irrelevant metrics from a large number of software metrics available. Prior studies \cite{Wang:A,Wang: A2,Liu:Toward,Peng: Feature} lay a solid foundation for our work.

\section{Problem and Approach}
\subsection{Analysis of Defect Prediction Problem}
\begin{figure*}
  \centering
  \includegraphics[width=5.5in,height=3.5in]{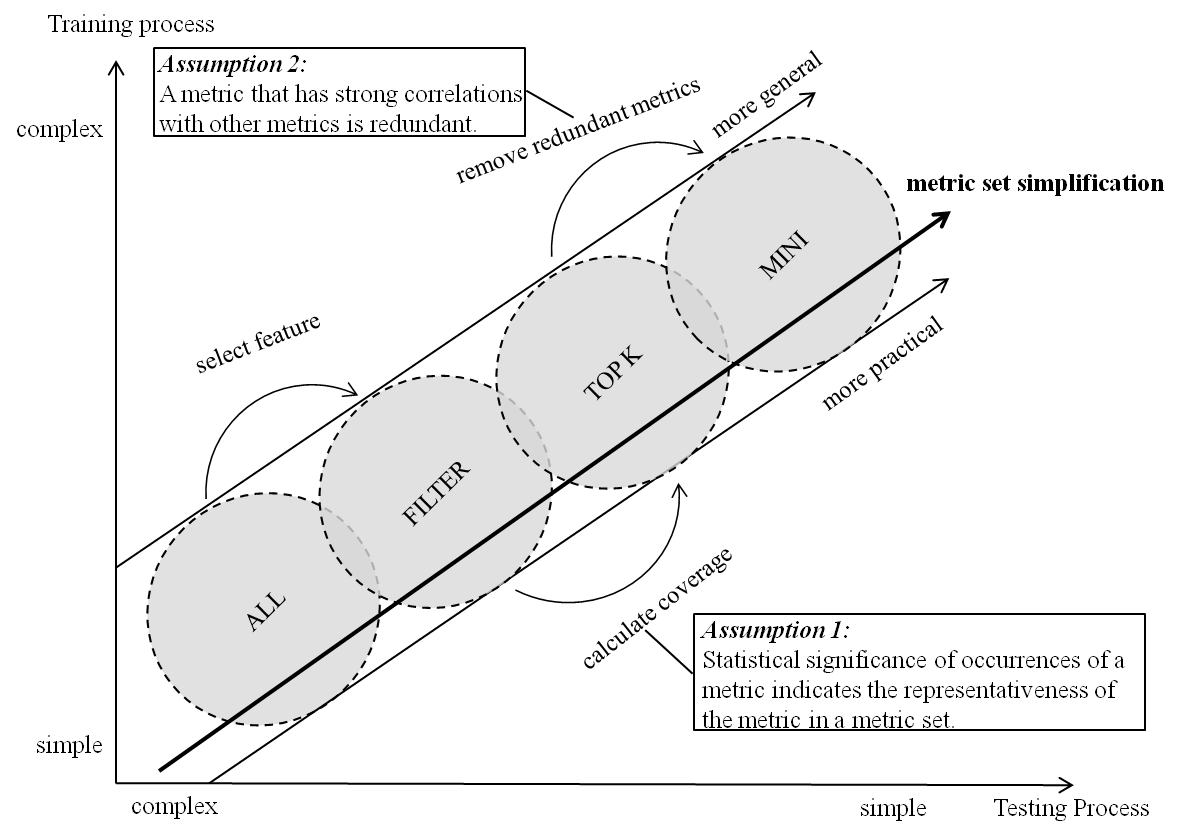}\\
  \caption{Simplified metric set building process.}\label{approach}
\end{figure*}

Machine learning techniques have emerged as an effective way to predict the existence of a bug in a change made to a source code file \cite{Shivaji: Reducing}. A classifier learns using training data, and it is then used for test data to predict bugs or bug proneness. During the learning process, one of the easiest methods is to directly train prediction models without the introduction of any attribute/feature selection techniques (see Figure \ref{approach}). However, this treatment will increase the burden on features analysis and the process of learning. Moreover, it is also easy to generate information redundancy and increase the complexity of the prediction models. For a large feature set, on one hand, the computation complexity of some feature values may be very high, and the cost of data acquisition and processing is far more than their contributions to a predictor; on the other hand, the addition of many useless or correlative features is harmful to a predictor's accuracy, and high complexity of a predictor will affect its generalization capability.

A reasonable method to deal with large feature set is to perform a feature selection process, so as to identify that a subset of features can provide the best classification result. Feature selection can be broadly classified as \emph{feature ranking} and \emph{feature subset selection}, or be categorized as \emph{filters} and \emph{wrappers}. \emph{Filters} are algorithms in which a subset of features is selected without involving any learning algorithm, whereas \emph{wrappers} are algorithms that use feedback from a classification learning algorithm to determine which feature(s) to include in constructing a classification model. In the literature \cite{Shivaji: Reducing,Wang:A,Wang: A2,Liu:Toward}, many approaches have been proposed to discard less important features in order to improve defect prediction. The more refined a feature subset becomes, the more stable a feature selection algorithm is.

Feature selection substantially reduces the number of features, and a reduced feature subset permits better and faster bug predictions. Despite this result, the stability of feature selection techniques depends largely on the context of defect prediction models. In other words, a feature selection technique can perform well in a data set, but perhaps, the effect will become insignificant when crossing other data sets. Moreover, the generality of the obtained feature subset is very poor.  To our knowledge, each successful prediction model in prior studies usually uses no more than 10 metrics \cite{Jureczko:Significance}. In this study, first, we record the number of occurrences of different metrics in each prediction model, and then use the Top-\emph{k} representative metrics as a universal feature subset to predict defect for all projects. This approach will be more suitable for projects without sufficient historical data for WPDP because of its generality. The greater the prediction models training, the more general the Top-\emph{k} feature subset is.

However, there are still some strong correlations among the Top-\emph{k} metrics obtained according to the number of occurrences. Hence, another desirable method is to minimize such a feature subset by applying some reduction criteria, for example, discarding the metric that has strong correlations with other metrics within the Top-\emph{k} subset. To the best of our knowledge, the simple static code metric such as LOC has been validated to be a useful predictor of software defects \cite{Zhang:An}. Furthermore, there is a sufficient amount of data available for these simple metrics to train any prediction models. Whether there is a simplified (or even minimum) feature subset that performs well both within a project and across projects as long as there is a sufficient number of training models. As depicted in Figure \ref{approach}, we define this progressive reduction on the size of feature set as \emph{metric set simplification}, which represents the primary contribution throughout our study.

\subsection{Research Questions}
According to the review of existing work related to  (1) the trade-off between WPDP and CPDP, and (2) the choice of software metrics and classifiers, we attempt to find empirical evidence that addresses the following four research questions in our paper:

\begin{itemize}
      \item \emph{RQ1: Which type of defect prediction models is more suitable for a specific project between WPDP and CPDP?}\\
      Prediction accuracy depends not only on the learning techniques applied, but also on the selected training data sets. It is common sense that training data obtained from the same project will perform better than those collected from other projects. To our surprise, He \emph{et al.} \cite{He:An} found the opposite result that the latter is better than the former. We thus further validate the hypothesis that the former will be more suitable when emphasizing the precision as a result of the authenticity of the data, and the latter in turn will be more preferable when emphasizing the recall  and F-measure with sufficient information.
     \item \emph{RQ2: Does the predictor built with a simplified metric set work well?}\\
      In practice, software engineers have to make a trade-off between accuracy and effort-and-cost in software quality control processes. There is no doubt that more effort-and-cost must be paid for data acquisition and processing when taking more metrics into account in a prediction model, although including more information may improve prediction accuracy to some extent. For this question, we would like to validate whether a predictor based on few representative metrics can still achieve acceptable prediction results. If so, the generality of the procedure for defect prediction will be remarkably improved.
  \item \emph{RQ3: Which classifier is more likely to be the choice of defect prediction with a simplified metric set?}\\
      Prior studies suggest that easy-to-use classifiers tend to perform well, such as Na\"{\i}ve Bayes and Logistic Regression \cite{Hall:A}. Does this conclusion still hold when using simplified metric set for defect prediction? In addition, is the stability of results obtained from our approach with different classifiers statistically significant?
  \item \emph{RQ4: Is there a minimum metric subset that facilitates the procedure for general defect prediction?}\\
      It is well-known that you cannot have your cake and eat it too. Eliminating strong correlations between the metrics within the Top-$k$ metric subset leads to more generality, which might result in a loss of precision. What we would like to discuss is the existence of a minimum metric subset, which facilitates the procedure of general defect prediction with regard to the practicable criteria for acceptable results. For instance, recall $>0.7$ and precision $>0.5$ \cite{He:An}.

\end{itemize}

\subsection{Simplification of Metric Set}
\subsubsection{Top-k Feature Subset }
As shown in Figure \ref{approach}, we constructed four combinations of software metrics as experimental subjects to carry out our experiments. \emph{ALL} indicates that no feature selection techniques are introduced when constructing defect prediction models in our experiments, and \emph{FILTER} indicates that a feature selection technique with a \emph{CfsSubsetEval} evaluator and \emph{GreedyStepwise} search algorithm in Weka\footnote{http://www.cs.waikato.ac.nz/ml/weka/} is introduced to select features from original data sets automatically. The \emph{CfsSubsetEval} evaluator evaluates the worth of a subset of attributes by considering the individual predictive ability of each feature along with the degree of redundancy between them. The \emph{GreedyStepwise} algorithm performs a greedy forward or backward search through the space of attribute subsets, which may start with no/all attributes or from an arbitrary point in the space and stops when the addition/deletion of any remaining attributes results in a decrease in the evaluation. It is worthwhile to note that \emph{filters} are usually less computationally intensive than \emph{wrappers}, which depend largely on a specific type of predictive models.

On the basis of feature selection techniques, we use \emph{TOPK} to represent the Top-\emph{k} metrics determined by the number of occurrences of different metrics in the obtained filtering models. To identify the optimal \emph{K} value of the \emph{TOPK} metric subset, we introduce a \emph{Coverage} index, which is used to measure the degree of coverage between two groups of metrics from the same data set (i.e., \emph{FILTER} vs. \emph{TOPK}). In this paper, we use the \emph{Coverage} index because it takes the representativeness of selected metrics into consideration. Suppose $Filter_{i}$ is the metric subset selected automatically from data set $i$ and $Top_{k}$ is the $k$ most frequently occurring metrics, we compute the \emph{Coverage} value between two groups of metrics as follows:

\begin{equation}\label{Eq.2}
 Coverage(k)=\frac{1}{n}\sum_{i}^{n}\frac{Filter_{i}\bigcap Top_{k}}{Filter_{i}\bigcup Top_{k}},
\end{equation}
where $n$ is the total number of data sets, and $0\leq Coverage(k)\leq 1$. It is proportional to the intersection between the top $k$ frequently occurring metrics and the subset of metrics obtained by the feature selection technique, and is inversely proportional to their union. If these two groups of metrics are the same, the measure is 1. The greater the measure becomes, the more representative the \emph{TOPK} is.

\begin{figure*}
  \centering
  \includegraphics[width=6in]{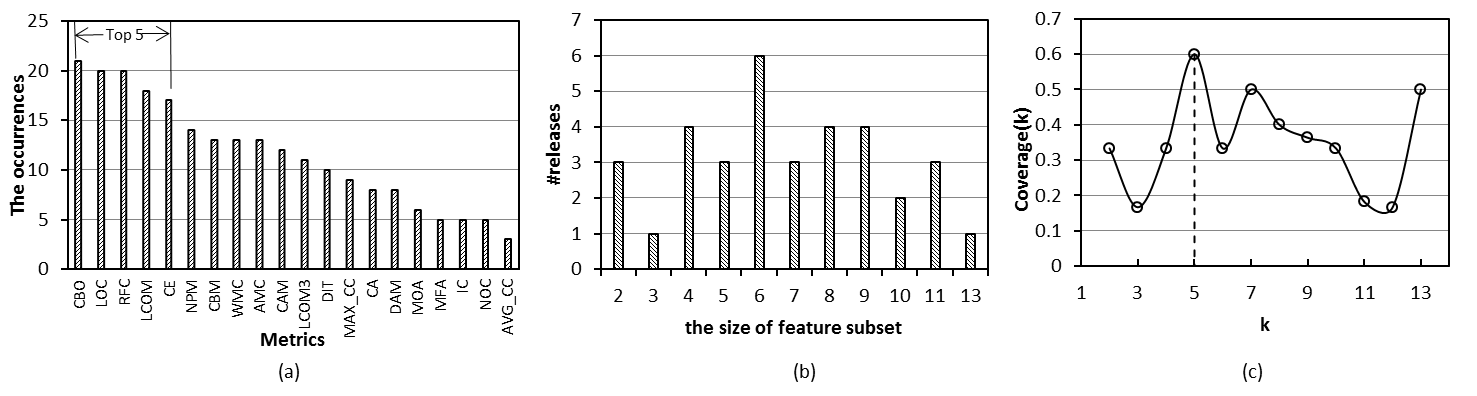}\\
  \caption{The determination of the optimal $k$ value.}\label{occurrence}
\end{figure*}

The range of parameter $k$ depends not only on the number of occurrences of each metric, but also on the size of the subset of the metrics selected with Weka. The optimal $k$ value is determined by the index \emph{Coverage} of each \emph{TOPK} combination. For example, for the data sets in this study, the number of occurrences of the top 5 metrics is more than 17 compared with the total number of occurrences 34: CBO (21), LOC (20), RFC (20), LCOM (18), and CE (17) (see Figure \ref{occurrence}(a)). Meanwhile, the feature subsets of most releases analyzed have no more than 10 metrics (see Figure \ref{occurrence}(b)), and the value of Coverage(5) reaches a peak (0.6) (see Figure \ref{occurrence}(c)).

\subsubsection{Minimum Feature Subset}
Although the \emph{TOPK} metric subset largely reduces the dimension of the original data, there are still strong correlations amongst the metrics within this subset. In order to alleviate information redundancy or remove redundant metrics, we further screen the metric set to determine the minimum subset by the following three setups:

\begin{enumerate}
  \item[(1)] Constructing all possible combinations $C$ among the Top-\emph{k} metrics, $C=\{C_1,C_2,\cdots, C_p\}\quad (p=2^k-1)$, and $C_h$=$\{m_1, m_2,\cdots, m_i\}\quad (m_i\in TOPK, i\leq k, h\leq p)$. Take the top 5 metrics as an example, $\{$CBO, LOC$\}$ is one of the $C^2_5$ combinations. For convenience, the $\{$CBO, LOC$\}$ combination is symbolized as CBO+LOC.
  \item[(2)] Calculating correlation coefficient matrix $R^{k\times k}$. If the element $r_{ij}$ in $R$ is larger than $\varphi$, excluding all combinations $C_h$ that include the metrics $m_i$ and $m_j$ from $C$, and returning the remaining metric subset $C' (|C'|\leq |C|)$.
  \item[(3)] Finally, determining the minimum metric subset by the \emph{Coverage} index. That is, replacing $Top_k$ with $C^{'}_h (C^{'}_h \in C')$ in Equation (\ref{Eq.2}).
\end{enumerate}

In setup (2), the element $r_{ij}$ is the correlation coefficient between the $i^{th}$ metric and the $j^{th}$ metric. In general, $r_{ij}>0$ indicates a positive correlation between two metrics, $r_{ij}<0$ indicates a negative correlation, whereas $r_{ij}=0$ indicates no correlation. The closer to $1$ the absolute value of $r$ is, the greater the strength of the correlation between metrics becomes. Although there is no clear boundary for the strength of correlations, as a rule of thumb, the guideline in Table \ref{range} is often useful. In our study, the greater the strength of the correlation between two metrics becomes, the more redundant the existing information is. For example, if the correlation coefficient $r$ between CBO and CE is larger than $\varphi$, all combinations that contain these two metrics have to be excluded, such as CBO+CE, CBO+CE+LOC, CBO+CE+RFC, and so on.

In setup (3), we compute the \emph{Coverage} values of remaining combinations again. Besides, we further validate the results of the minimum metric subset based on the corresponding thresholds of \emph{Recall}, \emph{Precision} and \emph{F-measure}. There is also no unified standard for judging whether the result of a defect prediction model is successful. Different studies may use different thresholds to evaluate their results. For example, Zimmermann \emph{et al.} \cite{Zimmermann:Cross} judged their results with all \emph{Recall}, \emph{Precision}, and \emph{Accuracy} values greater than 0.75. Nevertheless, He \emph{et al.} \cite{He:An} made predictions with \emph{Recall} greater than 0.7 and \emph{Precision} greater than 0.5 with regard to good engineering practices. Hence, the thresholds used rely on the previous studies of some other researchers and our own research experience on defect prediction.

\begin{table}\small
  \centering
  \caption{The strength of correlations.}\label{range}
  \begin{tabular}{|c|c|}
    \hline
    Absolute Value of $r$ & Strength \\ \hline
    0.8-1 & Very Strong \\ \hline
    0.6-0.8 & Strong \\ \hline
    0.4-0.6 & Moderate \\ \hline
    0.2-0.4 & Weak \\ \hline
    0.0-0.2 & None or very weak \\
    \hline
  \end{tabular}

\end{table}

\section{Experimental Setup}
\subsection{Data Collection}
In our study, 34 releases of 10 open-source projects available at the PROMISE repository are used for validation, a total of 34 different defect data sets. Detailed information on the 34 data sets is listed in Table \ref{data}, where $\#Instances$ and $\#Defects$ are the number of instances and the number of defects respectively. The last column is the ratio of buggy classes to all classes. Each instance in these public data sets represents a class file of a release and consists of two parts: independent variables including 20 static code metrics (e.g., CBO, WMC, RFC, LCOM, etc.) and a dependent variable labeling how many bugs are in this class. Table \ref{metrics} presents all of the variables involved in our study.

\begin{figure}
\centering
\includegraphics[width=3.5in,height=1.5in]{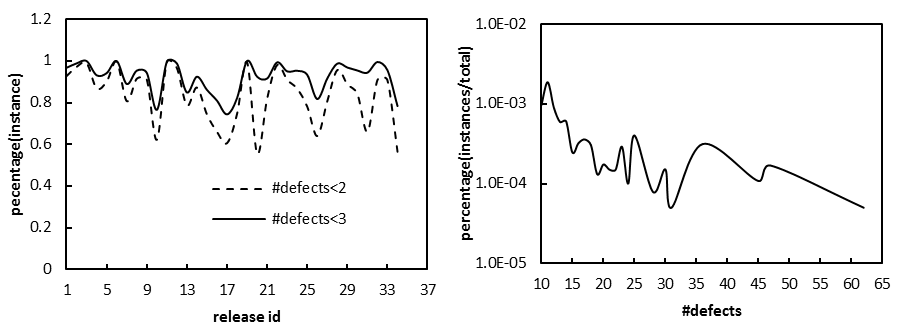}
\caption{The distribution of defects in the data sets.}
\label{defect}
\end{figure}

Note that, there is a preprocessing that transforms the bug attribute into a binary classification before using it as the dependent variable in our context. The reason why we use such a preprocessing consists in two regions. On one hand, the majority of class files in the 34 data sets have no more than 3 defects. On the other hand, the ratio of the instances with more than 10 defects to the total instances is less than 0.2\% (see Figure \ref{defect}). Furthermore, the preprocessing has been used in several prior researches, such as \cite{Peters:Better,He:An,Turhan:On,Turhan:Empirical,Herbold:Training}, to predict defect proneness. In a word, a class is non-buggy only if the number of bugs in it is equal to 0. Otherwise, it is buggy. A defect prediction model typically labels each class as either buggy or non-buggy.

\begin{table}\small
\centering
\caption{Details of the 34 data sets, including the number of instances (files), defects and defect-proneness.}\label{data}
\begin{tabular}{c|c|c|c|c}
  \hline
  No. & Releases  & \#Instances(Files) &\#Defects &\%Defects \\ \hline
  1  & Ant-1.3   &  125   &  20   & 16.0    \\
  2  & Ant-1.4   &  178   &  40   & 22.5 \\
  3  & Ant-1.5   &  293   &  32   & 10.9 \\
  4  & Ant-1.6   &  351   &  92   & 26.2 \\
  5  & Ant-1.7   &  745   &  166  & 22.3 \\
  6  & Camel-1.0 &  339   &  13   &  3.8   \\
  7  & Camel-1.2 &  608   &  216  &  35.5 \\
  8  & Camel-1.4 &  872   &  145  & 16.6 \\
  9  & Camel-1.6 &  965   &  188  &  19.5 \\
  10 & Ivy-1.1   &  111   &  63   &  56.8 \\
  11 & Ivy-1.4   &  241   &  16   &  6.6 \\
  12 & Ivy-2.0   &  352   &  40   &  11.4 \\
  13 & Jedit-3.2 &  272   &  90   &  33.1 \\
  14 & Jedit-4.0 &  306   &  75   &  24.5 \\
  15 & Lucene-2.0&  195   &  91   &  46.7 \\
  16 & Lucene-2.2&  247   &  144  &  58.3 \\
  17 & Lucene-2.4&  340   &  203  &  59.7 \\
  18 & Poi-1.5   &  237   &  141  &  59.5 \\
  19 & Poi-2.0   &  314   &  37   &  11.8 \\
  20 & Poi-2.5   &  385   &  248  &  64.4 \\
  21 & Poi-3.0   &  442   &  281  &  63.6 \\
  22 & Synapse-1.0 &  157   &  16   &  10.2 \\
  23 & Synapse-1.1 &  222   &  60   &  27.0 \\
  24 & Synapse-1.2 &  256   &  86   &  33.6 \\
  25 & Velocity-1.4&  196   &  147  &  75.0 \\
  26 & Velocity-1.5&  214   &  142  &  66.4 \\
  27 & Velocity-1.6&  229   &  78   &  34.1 \\
  28 & Xalan-2.4   &  723   &  110  &  15.2 \\
  29 & Xalan-2.5   &  803   &  387  &  48.2 \\
  30 & Xalan-2.6   &  885   &  411  &  46.4 \\
  31 & Xerces-init &  162   &  77   &  47.5 \\
  32 & Xerces-1.2  &  440   &  71   &  16.1 \\
  33 & Xerces-1.3  &  453   &  69   &  15.2 \\
  34 & Xerces-1.4  &  588   &  437  &  74.3 \\
  \hline
\end{tabular}
\end{table}

Given the skew distributions of the value of independent variables in most data sets, it is useful to apply a ``log-filter" to all numeric values \emph{v} with $ln(v)$ (to avoid numeric errors with $ln(0)$, all numbers $ln(v)$ are replaced with $ln(v+1))$ \cite{Menzies:Data,Song:A}. In addition, there are many other commonly-used methods in machine learning literature, such as min-max and $z$-score \cite{Nam:Transfer}.

\subsection{Experiment Design}

\begin{figure*}
\centering
\includegraphics[width=7in,height=4in]{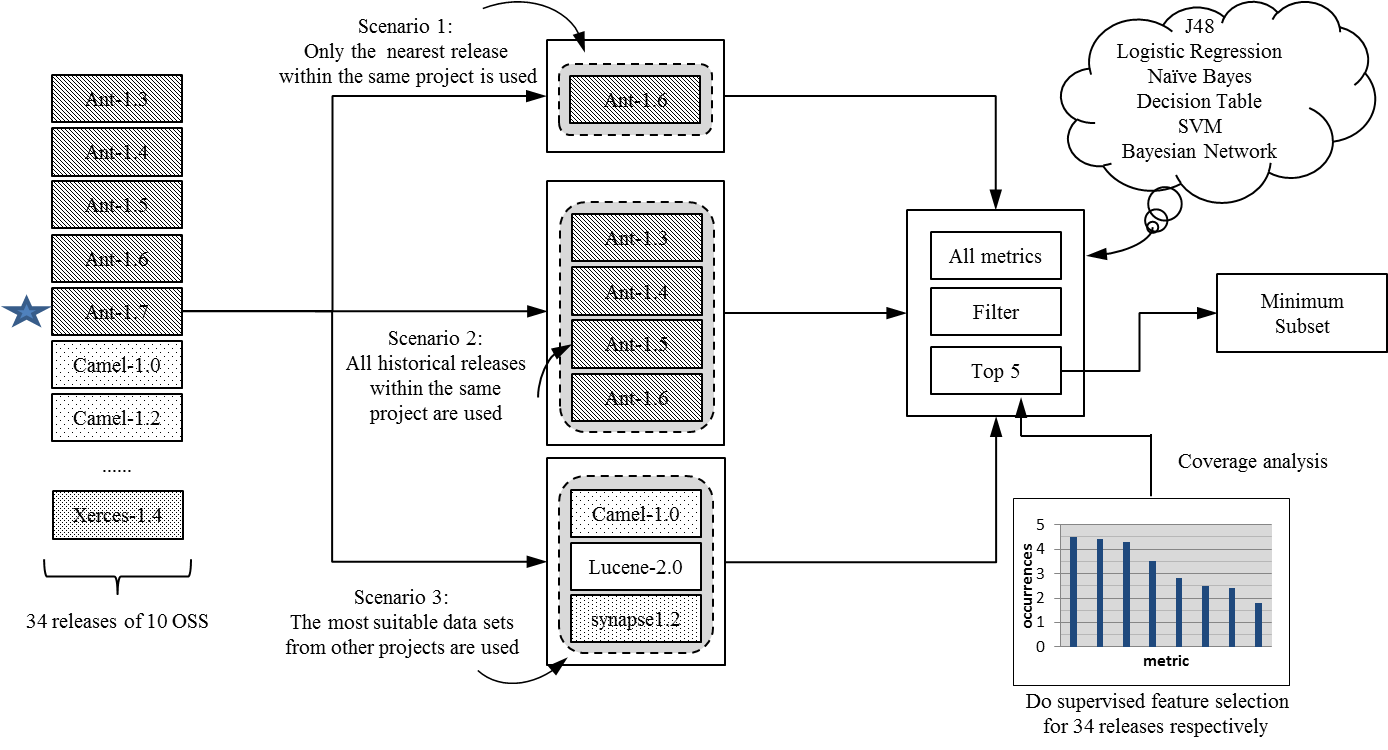}
\caption{The framework of our approach\textemdash an example of release Ant-1.7.}
\label{Fig.framework}
\end{figure*}

When you conduct an empirical study on defect prediction, many decisions have to be made regarding the collection of representative data sets, the choice of independent and dependent variables, modeling techniques, evaluation methods and evaluation criteria. The entire framework of our experiments is illustrated in  Figure \ref{Fig.framework}.

First, to make a comparison between WPDP and CPDP, three scenarios were considered in our experiments. (1) Scenario 1 (WPDP-1) uses the nearest release before the release in question as training data; (2) Scenario 2 (WPDP-2) uses all historical releases prior to the release in question as training data; (3) Scenario 3 (CPDP) selects the most suitable releases from other projects in terms of the method in \cite{He:An} as training data. For Scenario 1 and Scenario 2, the first release of each project is just used as training data for the upcoming releases. Thus, there are 24 ($34 - 10 = 24$) groups of tests among all the 34 releases of 10 projects. In order to ensure the comparability of experimental results of WPDP and CPDP, we selected 24 groups of corresponding tests for CPDP, though there is a total of 34 test data sets. For Scenario 3, the most suitable training data from other projects are generated by an exhaustive combinatorial test, rather than the three-step approach proposed in that paper. Specifically speaking, all the combinations of training data in their experiments consist of no more than three releases. For more details, please refer to the section 4.1 in \cite{He:An}.

Second, we applied six defect prediction models built with typical classifiers to 18 cases ($3 \times 6=18$), and compared the prediction results of three types of predictors based on different numbers of metrics.

Third, on the basis of the \emph{TOP5} metric subset, we further sought the minimum metric subset and tested the performance of the predictor built with such a minimum metric subset.

After this process is completed, we will discuss the answers to the four research questions of our study.

\subsection{Variables}
\textbf{Independent Variables.} The independent variables represent the inputs or causes, and are also known as predictor variables or features. It is usually what you think will affect the dependent variable. In our study, there are 20 commonly-used static code metrics, including CK suite (6), Martin's metrics (2), QMOOM suite (5), Extended CK suite (4), and McCabe's CC (2) as well as LOC. Each one exploits a different source of code information (see Table \ref{metrics}). For example, the value of the WMC is equal to the number of methods in a class (assuming unity weights for all methods) \cite{Jureczko:Towards}.

\begin{table}\small
\centering
\caption{Twenty independent variables of static code metrics and one dependent variable in the last row.}\label{metrics}
\begin{tabular}{cc} \hline
 \textbf{Variable}&\textbf{Description} \\ \hline
 \multicolumn{2}{c}{CK suite (6)}  \\ \hline
  WMC & Weighted Methods per Class  \\
  DIT & Depth of Inheritance Tree\\
  LCOM & Lack of Cohesion in Methods \\
  RFC & Response for a Class\\
  CBO & Coupling between Object classes \\
  NOC & Number of Children\\ \hline
\multicolumn{2}{c}{Martin¡¯s metric (2)}  \\ \hline
  CA & Afferent Couplings \\
  CE & Efferent Couplings\\ \hline
\multicolumn{2}{c}{QMOOM  suite (5)} \\ \hline
  DAM & Data Access Metric \\
  NPM & Number of Public Methods\\
  MFA & Measure of Functional Abstraction\\
  CAM & Cohesion Among Methods \\
  MOA & Measure Of Aggregation \\ \hline
\multicolumn{2}{c}{Extended CK suite (4)} \\ \hline
    IC  & Inheritance Coupling \\
    CBM & Coupling Between Methods\\
    AMC & Average Method Complexity \\
    LCOM3 & Normalized version of LCOM\\ \hline
\multicolumn{2}{c}{McCabe's CC (2)} \\ \hline
    MAX$\_$CC & Maximum values of methods in the same class \\
  AVG$\_$CC & Mean values of methods in the same class \\ \hline
  LOC & Lines Of Code \\ \hline
  Bug & non-buggy or buggy\\ \hline
\end{tabular}
\end{table}

\textbf{Dependent Variable.} The dependent variable represents the output or effect, and is also known as response variable. In mathematical modeling, the dependent variable is studied to see if and how much it varies as the independent variables vary. The goal of defect prediction in our experiments is to identify defect-prone classes precisely for a given release. We deem the defect-proneness as a binary classification problem (i.e., buggy vs. non-buggy). That is to say, a class is non-buggy only if the number of bugs in it is equal to 0; otherwise, it is buggy.

\subsection{Classifiers}
In general, an algorithm that implements classification, especially in a concrete implementation, is known as a classifier. There are inconsistent findings regarding the superiority of a particular classifier over others \cite{Lessmann: Benchmarking}. In this study, software defect prediction models are built with six well-known classification algorithms used in \cite{He:An}, namely, J48, Logistic Regression (LR), Na\"{\i}ve Bayes (NB), Decision Table (DT), Support Vector Machine (SVM) and Bayesian Network (BN). All classifiers were implemented in Weka. For our experiments, we used the default parameter settings for different classifiers specified in Weka unless otherwise specified.

\textbf{J48} is an open source Java implementation of the C4.5 decision tree algorithm. It uses the greedy technique for classification and generates decision trees, the nodes of which evaluate the existence or significance of individual features. Leaves in the tree structure represent classifications and branches represent judging rules.

\textbf{Na\"{\i}ve Bayes (NB)} is one of the simplest classifier based on conditional probability. The classifier is termed as ``na\"{\i}ve" because it assumes that features are independent. Although the independence assumption is often violated in the real world, the Na\"{\i}ve Bayes classifier often competes well with more sophisticated classifiers \cite{Rish: An}. The prediction model constructed by this classifier is a set of probabilities. Given a new class, the classifier estimates the probability that the class is buggy based on the product of the individual conditional probabilities for the feature values in the class.

\textbf{Logistic Regression (LR)} is a type of probabilistic statistical regression model for categorical prediction by fitting data to a logistic curve \cite{Bishop: Pattern}. It is also used to predict a binary response from a binary predictor, used for predicting the outcome of a categorical dependent variable based on one or more features. Here, it is suitable for solving the problem in which the dependent variable is binary, that is to say, either buggy or non-buggy.

\textbf{Decision Table (DT)}, as a hypothesis space for supervised learning algorithm, is one of the simplest hypothesis spaces possible, and it is usually easy to understand \cite{Ron:The}. A decision table has two components: a schema, which is a set of features, and a body, which is a set of labeled instances.

\textbf{Support Vector Machine (SVM)} is a supervised learning model with associated learning algorithms that is typically used for classification and regression analysis by finding the optimal hyper-plane that maximally separates samples in two different classes. A prior study conducted by Lessmann \emph{et al.} \cite{Lessmann: Benchmarking} showed that the SVM classifier performed equally with the Na\"{\i}ve Bayes classifier in the context of defect prediction.

\textbf{Bayesian Network (BN)} is a graphical representation that presents the probabilistic causal or influential relationships among a set of variables of interest. Because BNs can model the intra-relationship between software metrics and allow one to learn about causal relationships, the BN learning algorithm is also a comparative candidate for building prediction models for software defects. For more details, please refer to \cite{Okutan:Software}.

\subsection{Evaluation Measures}
In this study, we used a binary classification technique to predict classes that are likely to have defects. A binary classifier can make two possible errors: \emph{false positives (FP)} and \emph{false negatives (FN)}. In addition, a correctly classified buggy class is a \emph{true positive (TP)} and a correctly classified non-buggy class is a \emph{true negative (TN)}. We evaluated binary classification results in terms of \emph{Precision}, \emph{Recall}, and \emph{F-measure}, which are described as follows:
\begin{itemize}
  \item \emph{Precision} addresses how many of the classes returned by a model are actually defect-prone. The best precision value is 1. The higher the precision is, the fewer false positives (i.e., non-defective elements incorrectly classified as defect-prone one) exist:
    \begin{equation}\label{Eq.precision}
        Precision=\frac{TP}{TP+FP}.
    \end{equation}
  \item \emph{Recall} addresses how many of the defect-prone classes are actually returned by a model. The best recall value is 1. The higher the recall is, the lower the number of false negatives (i.e., defective classes missed by the model) is:
     \begin{equation}\label{Eq.recall}
        Recall=\frac{TP}{TP+FN}.
     \end{equation}
  \item \emph{F-measure} considers both \emph{Precision} and \emph{Recall} to compute the accuracy, which can be interpreted as a weighted average of \emph{Precision} and \emph{Recall}. The value of \emph{F-measure} ranges between 0 and 1, with values closer to 1 indicating better performance for classification results.
     \begin{equation}\label{Eq.F}
       F-measure=\frac{2*Precise*Recall}{Precise+Recall}.
     \end{equation}
\end{itemize}

To answer \emph{RQ4} in the following sections, we also introduced a \emph{Consistency} index to measure the degree of stability of the prediction results generated by the predictors in question.

\begin{equation}\label{Eq.3}
 Consistency=\frac{dn-k^{2}}{k(n-k)}  (Consistency \leq 1),
\end{equation}
where $d$ is the number of actually defect-prone classes returned by a model in each data set ($d=TP$ ); $k$ is the total of actually defect-prone classes in the data set ($k=TP+FN$ ); $n$ is the total number of instances. If \emph{TP = TP + FN}, the \emph{Consistency} value is 1. The greater the \emph{Consistency} index becomes, the more stable a model is. Note that, Equation (\ref{Eq.3}) is often used to measure the stability of feature selection algorithms \cite{Kuncheva:A}. We introduced this equation to our experiments because of the same implication.


\section{Experimental Results}

In this section, we report the primary results so as to answer the four research questions formulated in Section 3.2.

\subsection{RQ1: Which type of defect prediction models is more suitable for a specific project between WPDP and CPDP?}

\begin{figure*}
\centering
\includegraphics[width=6.6in,height=7.7in]{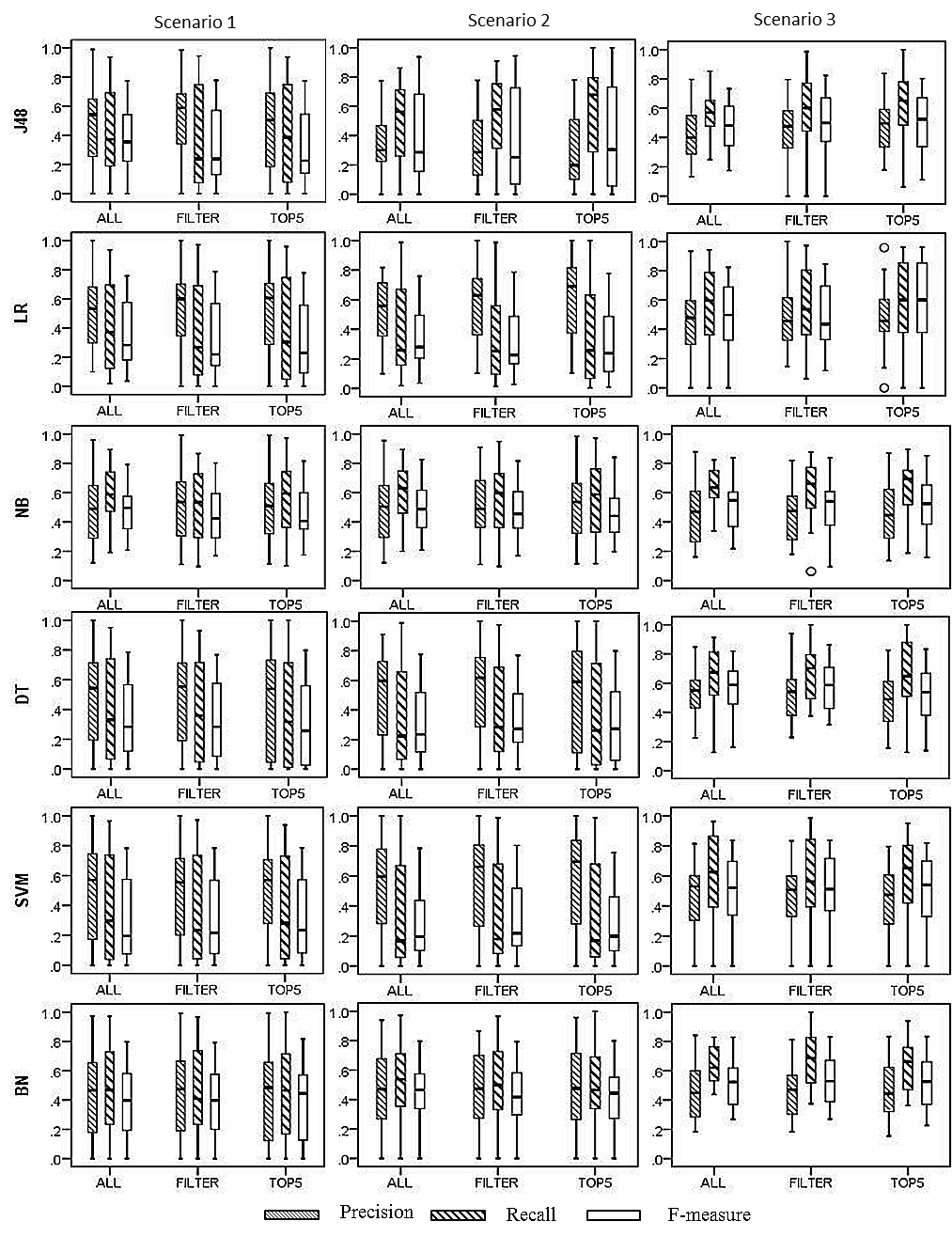}
\caption{The standardized boxplots of the performance achieved by different predictors based on J48, Logistic Regression, Na\"{\i}ve Bayes, \protect\\ Decision Table, SVM and Bayesian Network in different scenarios, respectively. From the bottom to the top of a standardized box plot: minimum, first quartile, median, third quartile, and maximum. Any data not included between the whiskers is plotted as a small circle.}
\label{Fig.2}
\end{figure*}

For each prediction model, Figure \ref{Fig.2} shows some interesting results. (1) The precision becomes higher when using training data within the same project, whereas CPDP models receive high recall and F-measure in terms of median value, implying a significant improvement in the accuracy. For example, for the \emph{TOP5} metric subset, in Scenario 1, the median precision, recall and F-measure of J48 are 0.504, 0.385 and 0.226, respectively, and in Scenario 3, they are 0.496, 0.651 and 0.526, respectively. (2) For WPDP, there is no significant difference in the accuracy between Scenario 1 and Scenario 2. For example, for the \emph{TOP5} metric subset, Scenario 1 vs. Scenario 2 turn out to be relatively matched in terms of the above measures for J48 (i.e., 0.504 vs. 0.679, 0.385 vs. 0.304, and 0.226 vs. 0.20). That is, the quantities of training data do not affect prediction results remarkably.

WPDP models generally capture higher precision than CPDP models, which, in turn, achieve high recall for two reasons. First, training data from the same project represent the authenticity of the project data and can achieve a higher precision based on historical data. Second, existing release sets of other projects may be not comprehensive enough to represent the global characteristics of the target project. In other words, training data from other projects may be more preferable because the rich information in the labeled data leads to the identification of more actually defect-prone classes. As opposed to our expectation, there is no observable improvement in the accuracy when increasing the number of training datasets in WPDP. This happens because the values of some metrics are identical among different releases. As shown in Figure \ref{Fig.2}, increasing the quantity of training data does not work better and even reduces the recall because of information redundancy.

The results also validate the idea that CPDP may be feasible for a project with insufficient local data. In addition, Figure \ref{Fig.2} shows a tendency that the \emph{TOP5} metric subset simplified by our approach appears to provide a comparable result to the other two cases. However, until now, we just analyzed the comparison of training data between WPDP and CPDP with six prediction models, without examining whether the predictor with a simplified metric set works well. For example, it might be possible that a predictor built with few metrics ( e.g., \emph{TOPK}) can provide a satisfactory prediction result with the merits of less effort-and-cost. This analysis is the core of our work and will be investigated in the upcoming research questions.

\begin{table*}\small
\centering
\caption{The performance of the predictors built with \emph{ALL} and \emph{TOP5} metrics and the comparison of the distributions of their results in \protect\\terms of the Wilcoxon signed-rank test and Cliff's effect size ($d$): the underlined numbers in the Median \emph{TOP5/ALL} column represent that \protect\\ the result of \emph{TOP5} is better, and those in the \emph{ALL} vs. \emph{TOP5} column represent that one can reject the \emph{null hypothesis}; the negative numbers in bold represent that the result of \emph{TOP5} is better. }\label{Simple1}
\begin{tabular}{c|c|c|c|c|c|c|c|c|c|c} \hline
 \multicolumn{2}{c|}{\multirow{2}*{ }}
 &\multicolumn{3}{c|}{ Median \emph{ALL} value} &\multicolumn{3}{c|}{Median \emph{TOP5}/\emph{ALL} } &\multicolumn{3}{c}{\emph{ALL} vs. \emph{TOP5} ($Sig.p <0.01$ ($d$))}
 \\
 \cline{3-11}
 \multicolumn{2}{c|}{}
  & \multicolumn{1}{c|}{Precision} &\multicolumn{1}{c|}{Recall} &\multicolumn{1}{c|}{F-measure}
  & \multicolumn{1}{c|}{Precision} &\multicolumn{1}{c|}{Recall} &\multicolumn{1}{c|}{F-measure}
  & \multicolumn{1}{c|}{Precision} &\multicolumn{1}{c|}{Recall} &\multicolumn{1}{c}{F-measure}
   \\ \hline
\multirow{6}*{Scen. 1}
    &J48  & 0.543 & 0.372 & 0.355 & 0.928 & \underline{1.04} & 0.638 & 0.485(\textbf{-0.003}) & 0.548(0.083) & 0.058(0.142)\\
    &LR   & 0.532 & 0.372 & 0.282 & \underline{1.14} & 0.815 & 0.808 & 0.465(\textbf{-0.009}) & 0.089(0.106) & 0.015(0.142) \\
    &NB   & 0.490 & 0.588 & 0.496 & \underline{1.04} & \underline{1.01}  & 0.819 & \underline{0.009}(\textbf{-0.066}) & 0.092(0.063) & 0.71(0.031)\\
    &DT   & 0.543 & 0.331 & 0.284 & 0.992 & 0.960 & 0.904 & 0.959(0.009) & 0.076(0.095) & 0.022(0.083) \\
    &SVM  & 0.571 & 0.298 & 0.197 & 0.999 & 0.943 & \underline{1.43} & 0.546(0.016) & 0.709(\textbf{-0.016}) & 0.289(\textbf{-0.049}) \\
    &BN  & 0.467 & 0.470 & 0.397 & \underline{1.04} & 0.992 & \underline{1.12} & 0.765(0.031) & 0.573(\textbf{-0.012}) & 0.575(0.023) \\  \hline
\multirow{6}*{Scen. 2}
  &J48    & 0.564 & 0.287 & 0.301 & \underline{1.20} & \underline{1.06} & 0.663 & 0.064(\textbf{-0.149}) & 0.563(0.12) & 0.114(0.189)\\
  &LR     &0.557 & 0.259 & 0.280 & \underline{1.24}  & 0.993 & 0.853 & 0.024(\textbf{-0.257}) & 0.116(0.132) & 0.021(0.177)\\
  &NB     &0.504 & 0.629 & 0.487 & \underline{1.06}  & 0.934 & 0.906 &\underline{ 0.002}(\textbf{-0.118})& \underline{0.005}(0.142)& 0.116(0.066)\\
  &DT    &0.597 & 0.224 & 0.235 & 0.988 & \underline{1.17} & \underline{1.16} & 0.520(0.007) & 0.936(0.0) & 0.421(\textbf{-0.009})\\
  &SVM  & 0.596 & 0.168 & 0.197 & \underline{1.17} & \underline{1.01} & \underline{1.01} & 0.279(\textbf{-0.08}) & 0.639(0.026) & 0.685(0.021) \\
  &BN  & 0.471 & 0.537 & 0.466 & \underline{1.01} & 0.868 & 0.953 & 0.305(\textbf{-0.026}) & 0.370(0.042) & 0.244(0.068) \\  \hline
\multirow{6}*{Scen. 3}
    &J48  & 0.399 & 0.570 & 0.482 & \underline{1.24} & \underline{1.14} & \underline{1.09} & 0.037(\textbf{-0.122})& 0.137(\textbf{-0.224}) & 0.103(\textbf{-0.08}) \\
    &LR   & 0.457 & 0.538 & 0.431 & \underline{1.0} & \underline{1.11} & \underline{1.19} & 0.738(\textbf{-0.033}) & 0.295(\textbf{-0.092}) & 0.831(\textbf{-0.012})\\
    &NB   & 0.471 & 0.634 & 0.548 & 0.948 & \underline{1.09} & 0.957 & 0.179(0.021)& 0.304(\textbf{-0.09}) & 0.475(\textbf{-0.01})\\
    &DT   & 0.549 & 0.674 & 0.558 & 0.891 & 0.959 & 0.916 & \underline{0.001}(0.177) & 0.761(\textbf{-0.009}) & 0.016(0.13)\\
    &SVM  & 0.530 & 0.626 & 0.521 & 0.895 & \underline{1.04} & \underline{1.04} & 0.153(0.040) & 0.498(0.047) & 0.136(0.054)\\
    &BN  & 0.449 & 0.621 & 0.524 & 0.986 & \underline{1.07} & \underline{1.01} & 0.290(\textbf{-0.024}) & 0.951(\textbf{-0.009}) & 0.137(\textbf{-0.028}) \\ \hline
\end{tabular}
\end{table*}

\begin{table*}\small
\centering
\caption{The performance of the predictors built with \emph{FILTER} and \emph{TOP5} metrics and the comparison of the distributions of their results in \protect\\ terms of the Wilcoxon signed-rank test and Cliff's effect size ($d$): the underlined numbers and the negative numbers in bold represent the same meaning as in Table \ref{Simple1}.}\label{Simple2}
\begin{tabular}{c|c|c|c|c|c|c|c|c|c|c} \hline
 \multicolumn{2}{c|}{\multirow{2}*{ }}
 &\multicolumn{3}{c|}{Median \emph{FILTER} value} &\multicolumn{3}{c|}{Median \emph{TOP5}/\emph{FILTER} } &\multicolumn{3}{c}{\emph{FILTER} vs. \emph{TOP5} ($Sig.p <0.01$ ($d$))}
 \\
 \cline{3-11}
 \multicolumn{2}{c|}{}
  & \multicolumn{1}{c|}{Precision} &\multicolumn{1}{c|}{Recall} &\multicolumn{1}{c|}{F-measure}
  & \multicolumn{1}{c|}{Precision} &\multicolumn{1}{c|}{Recall} &\multicolumn{1}{c|}{F-measure}
  & \multicolumn{1}{c|}{Precision} &\multicolumn{1}{c|}{Recall} &\multicolumn{1}{c}{F-measure}
   \\ \hline
\multirow{6}*{Scen. 1}
    &J48  & 0.588 & 0.238 & 0.238 & 0.856 & \underline{1.62}  & 0.953 & 0.267(0.085) & 0.306(\textbf{-0.036}) & 0.983(0.016)\\
    &LR   & 0.6 & 0.268 & 0.219 & \underline{1.01} & \underline{1.13} & \underline{1.04} & 0.664(0.017) & 0.823(0.005) & 0.362(0.042) \\
    &NB   & 0.536 & 0.535 & 0.424 & 0.95 & \underline{1.11}  & 0.959 & 0.563(0.04) & 0.277(\textbf{-0.049}) & 0.503(\textbf{-0.024})\\
    &DT   & 0.554 & 0.360 & 0.284 & 0.973 & 0.884 & 0.904 & 0.948(0.007) & 0.136(0.064) & 0.03(0.086) \\
    &SVM  & 0.555 & 0.234 & 0.218 & \underline{1.03} & \underline{1.20} & \underline{1.29} & 0.911(\textbf{-0.019}) & 0.809(\textbf{-0.010}) & 0.881(\textbf{-0.009})\\
    &BN  & 0.475 & 0.408 & 0.397 & \underline{1.02} & \underline{1.14} & \underline{1.12} & 0.881(0.024) & 0.845(\textbf{-0.035}) & 0.765(\textbf{-0.014}) \\   \hline
\multirow{6}*{Scen. 2}
  &J48    & 0.578 & 0.253 & 0.283 & \underline{1.17}  & \underline{1.20}  & 0.697 & 0.446(\textbf{-0.054}) & 0.647(0.014) & 0.248(0.056)\\
  &LR     & 0.630 & 0.253 & 0.226 & \underline{1.09}  & \underline{1.02} & \underline{1.06} & 0.449(\textbf{-0.09}) & 0.476(0.04) & 0.199(0.09)\\
  &NB     & 0.489 & 0.597 & 0.454 & \underline{1.097}  & 0.984 & 0.973 & 0.91(0.01)& 0.455(0.003) & 0.306(0.047)\\
  &DT     & 0.617 & 0.284 & 0.272 & 0.955 & 0.924 & \underline{1.0} & 0.717(0.063) & 0.717(0.031) & 0.198(0.052)\\
  &SVM  & 0.662 & 0.180 & 0.218 & \underline{1.05} & 0.941 & 0.916 & 0.741(\textbf{-0.076}) & 0.042(0.054) & 0.012(0.075)\\
  &BN  & 0.475 & 0.499 & 0.418 & \underline{1.01} & 0.933 & \underline{1.06} & 0.322(0.009) & 0.566(0.007) & 0.987(0.043) \\    \hline
\multirow{6}*{Scen. 3}
    &J48   & 0.476 & 0.601 & 0.499 & \underline{1.04} & \underline{1.08} & \underline{1.05} & 0.44(\textbf{-0.08})& 0.679(\textbf{-0.08}) & 0.819(\textbf{0.0}) \\
    &LR    & 0.473 & 0.594 & 0.488 & 0.966 & \underline{1.01} & \underline{1.05} & 0.831(\textbf{-0.02}) & 0.338(\textbf{-0.071}) & 0.927(\textbf{-0.003})\\
    &NB   & 0.474 & 0.66 & 0.54 & 0.941 & \underline{1.05} & 0.972 & 0.819(0.01)& 0.449(\textbf{-0.009}) & 0.265(\textbf{-0.014})\\
    &DT   & 0.549 & 0.702 & 0.588 & 0.891 & 0.921 & 0.916 & \underline{0.007}(0.151) & 0.592(0.036) & 0.016(0.127)\\
    &SVM  & 0.507 & 0.564 & 0.512 & 0.936 & \underline{1.16} & \underline{1.06} & \underline{0.003}(0.069) & 0.758(\textbf{-0.038}) & 0.058(0.049)\\
    &BN  & 0.470 & 0.686 & 0.529 & 0.942 & 0.965 & 0.996& 0.493(\textbf{-0.028}) & 0.076(0.139) & 0.493(0.026) \\   \hline
\end{tabular}
\end{table*}

\subsection{RQ2: Does the predictor built with a simplified metric set work well?}

\subsubsection{The Definition of Acceptable Result}

The balance of defect prediction between a desire for accuracy and generalization capability is an open challenge. The generalization capability of a defect prediction model is deemed a primary factor of prediction efficiency, while the accuracy achieved by a defect prediction model is a critical determinant of prediction quality. The trade-off between efficiency and accuracy requires an overall consideration of multiple factors. Hence, we defined two hypotheses as the acceptable condition in our study: on one hand, the results of a predictor based on few metrics are not worse than a benchmark predictor, or the overall performance ratio of the former to the latter is greater than 0.9 (in such a case, the overall performance of a predictor is calculated in terms of median value of evaluation measures); on the other hand, the distributions of their results have no statistically significant difference. We believe that the value of such a threshold can be acceptable according to software engineering practices.

Concerning the prediction results of the six prediction models in different contexts, Figure \ref{Fig.2} roughly suggests that such a simplified method can provide an acceptable prediction result compared with those more complex ones, especially in Scenario 3. For example, in Scenario 3, for the Na\"{\i}ve Bayes classifier with the top five metrics, the median precision, recall and F-measure values are 0.446, 0.694 and 0.525, respectively. Moreover, their maximal values of the three metric sets with the Na\"{\i}ve Bayes classifier are just 0.474, 0.694 and 0.548, respectively.

\subsubsection{The Comparison of Prediction Results }

To further validate the preferable prediction performance and practicability of the predictor built with a simplified metric set described above, we compared the performance of \emph{TOP5} against both \emph{ALL} and \emph{FILTER} in terms of a ratio of the former's median to the latter's median. The comparisons between \emph{TOP5} and \emph{ALL} shown in Table \ref{Simple1} (see the Median \emph{TOP5/ALL} column) present that more than 80 percent of the ratios are greater than 0.9, and some of them are labeled with an underline because their values are greater than 1. These results indicate that\textemdash compared with complex defect prediction models\textemdash the predictor built with the five frequently-used metrics in our data sets can achieve an acceptable result with little loss of precision. Similarly, the comparisons between \emph{TOP5} and \emph{FILTER} in Table \ref{Simple2} (see the Median \emph{TOP5/FILTER} column) also present an acceptable result based on the same evidence. However, we have to admit that the prediction results of different classifiers with the \emph{TOP5} metric subset in both Scenario 1 and Scenario 2 have several unacceptable cases, for example, the values of  \emph{F-measure} for J48 are under 0.7.

In addition to CPDP, WPDP using a simplified metric set (i.e., \emph{TOP5}) is still able to achieve a relatively high median precision (not less than 0.5, see Table \ref{Simple1} and Table \ref{Simple2}). The recall and F-measure for different predictors are stable except the Na\"{\i}ve Bayes model, which shows a sharp improvement in these two measures (see Figure \ref{Fig.2}). The significant increase indicates that more defect-prone classes can be identified by the Na\"{\i}ve Bayes learning algorithm. Therefore, the Na\"{\i}ve Bayes model appears to be more suitable for defect prediction when an engineer wants to use few metrics.

Figure \ref{Fig.2} only shows the standardized boxplots of the prediction results. In Table \ref{Simple1} and Table \ref{Simple2}, the performance of the predictor built with a simplified metric set is examined, whereas the last three columns of these two tables show the results of the Wilcoxon signed-rank test (\emph{p}-value) and Cliff's effect size ($d$) (i.e., $d$ is negative if the right-hand measure is higher than the left-hand one) \cite{Macbeth: Cliff}. Based on the \emph{null hypothesis} that two samples are drawn from the same distribution (i.e., $\mu_1-\mu_2=0$), the test is executed with an alternative hypothesis $\mu_1\neq \mu_2$.  The test yields a \emph{p}-value used to reject the \emph{null hypothesis} in favor of the alternative hypothesis. If the \emph{p}-value is more than 0.01 (i.e., there is no significant difference between the predictors under discussion), one cannot reject the \emph{null hypothesis} that both samples are in fact drawn from the same distribution. In our study, we considered the results of \emph{TOP5} as a target result, and thus, the statistical analyses were performed for \emph{ALL} vs. \emph{TOP5} and \emph{FILTER} vs. \emph{TOP5}.

In Table \ref{Simple1}, the Wilcoxon signed-rank test highlights that there are no significant differences between \emph{ALL} and \emph{TOP5}, indicated by the majority of $p > 0.01$ for the classifiers evaluated with the three measures, although four exceptions exist in Scenario 1 and Scenario 2. Additionally, note that for the effect size $d$, the predictor built with the \emph{TOP5} metric subset appears to be the choice that is more suitable for CPDP, as it is the only one that achieves the largest number of negative $d$ for different classifiers with ``no significant difference". Compared with \emph{ALL} metrics, the simplified metric set (i.e., \emph{TOP5}) achieves an improvement in \emph{Precision} for WPDP and an improvement in \emph{Recall} or \emph{F-measure} for CPDP. In short, with respect to the cost of computing twenty metrics, the simplified approach (e.g., \emph{TOP5} (25\% efforts)) is more practical under the specified conditions.

In Table \ref{Simple2}, there are also no significant differences between \emph{FILTER} and \emph{TOP5}, as indicated by the same evidence. Only two cases of \emph{Precision} in Scenario 3 present $p < 0.01$ when using Decision Table and SVM. The predictor built with the \emph{TOP5} metric subset also appears to be the choice that is more suitable for CPDP because of the majority of negative  $d$ for different classifiers. However, the improvement in \emph{Precision} for WPDP is less significant in Table \ref{Simple2}, but the improvement in \emph{Recall} or \emph{F-measure} is still supported for CPDP.

\subsubsection{Comparison with Existing Approaches}

To evaluate the usefulness of the proposed simplified approach, we built defect prediction models using two existing feature selection approaches (i.e., max-relevance (\emph{MaxRel}) and minimal-redundancy-maximal-relevance (\emph{mRMR}) \cite{Peng: Feature}) and performed experiments on all data sets in question. Then, we compared the results of our approach with the related methods according to the evaluation method used in the subsection 5.2.2.

In Table \ref{MaxRel}, the majority of values greater than 0.9 in the median \emph{TOP5/MaxRel} column present their comparative performance between \emph{TOP5} and \emph{MaxRel}. Furthermore, for WPDP, it is clear that there is an improvement in the precision by our approach, as a result of the majority of underlined median \emph{TOP5/MaxRel} values and negative $d$ values. For CPDP, our approach is obviously better than \emph{MaxRel} in terms of \emph{Recall}. What is more, compared with \emph{mRMR} in Table \ref{mRMR}, the advantage of our approach is especially obvious according to the values in the median \emph{TOP5/mRMR} column and the \emph{mRMR} vs. \emph{TOP5} column. On one hand, all the ratios of the three measures are larger than 0.9 and most of them are larger than 1. On the other hand, in three scenarios, additional evidence is the large number of negative $d$ values besides the majority of $p>0.01$ for the classifiers evaluated with the three measures.

With the evidence provided by the above activities, the proposed simplified approach is validated to be suitable for both WPDP and CPDP. We will further discuss the effectiveness of different classifiers and whether the existence of the minimum subset of metrics is substantiated.

\begin{table*}\small
\centering
\caption{The performance of the predictors built with \emph{MaxRel} and \emph{TOP5} metrics and the comparison of the distributions of their results in \protect\\ terms of the Wilcoxon signed-rank test and Cliff's effect size ($d$): the underlined numbers and the negative numbers in bold represent the same meaning as in Table \ref{Simple1}.}\label{MaxRel}
\begin{tabular}{c|c|c|c|c|c|c|c|c|c|c} \hline
 \multicolumn{2}{c|}{\multirow{2}*{ }}
 &\multicolumn{3}{c|}{Median \emph{MaxRel} value} &\multicolumn{3}{c|}{Median \emph{TOP5}/\emph{MaxRel} } &\multicolumn{3}{c}{\emph{MaxRel} vs. \emph{TOP5} ($Sig.p <0.01$ ($d$))}
 \\
 \cline{3-11}
 \multicolumn{2}{c|}{}
  & \multicolumn{1}{c|}{Precision} &\multicolumn{1}{c|}{Recall} &\multicolumn{1}{c|}{F-measure}
  & \multicolumn{1}{c|}{Precision} &\multicolumn{1}{c|}{Recall} &\multicolumn{1}{c|}{F-measure}
  & \multicolumn{1}{c|}{Precision} &\multicolumn{1}{c|}{Recall} &\multicolumn{1}{c}{F-measure}
   \\ \hline
\multirow{6}*{Scen. 1}
    &J48  & 0.454 & 0.339 & 0.246 & \underline{1.108} & \underline{1.137}  & 0.919 & 0.267(\textbf{-0.089}) & 0.068(\textbf{-0.069}) & 0.215(\textbf{-0.071})\\
    &LR   & 0.657 & 0.289 & 0.276 & 0.921 & \underline{1.050} & 0.828 & 0.370(0.101) & 0.748(\textbf{-0.021}) & 0.249(0.033) \\
    &NB   & 0.573 & 0.528 & 0.424 & 0.888 & \underline{1.126}  & 0.960 & 0.039(\textbf{-0.052}) & 0.709(\textbf{-0.021}) & 0.627(0.002)\\
    &DT   & 0.450 & 0.360 & 0.284 & \underline{1.198} & 0.883 & 0.904 & 0.619(\textbf{-0.052}) & 0.149(0.075) & 0.049(0.063) \\
    &SVM  & 0.549 & 0.264 & 0.208 & \underline{1.037} & \underline{1.065} & \underline{1.352} & 0.913(\textbf{-0.023}) & 0.616(\textbf{-0.049}) & 0.099(\textbf{-0.095})\\
    &BN  & 0.465 & 0.478 & 0.418 & \underline{1.044} & 0.976 & \underline{1.064} & 0.085(\textbf{-0.010}) & 0.210(0.021) & 0.586(0.017) \\   \hline
\multirow{6}*{Scen. 2}
  &J48    & 0.522 & 0.237 & 0.207 & \underline{1.300}  & \underline{1.285}  & 0.966 & 0.044(\textbf{-0.141}) & 0.500(0.016) & 0.913(0.002)\\
  &LR     & 0.677 & 0.240 & 0.224 & \underline{1.019}  & \underline{1.069} & \underline{1.066} & 0.903(\textbf{-0.002}) & 0.555(0.030) & 0.089(0.054)\\
  &NB     & 0.519 & 0.663 & 0.430 & \underline{1.033}  & 0.887 & \underline{1.027} & 0.089(\textbf{-0.040})& 0.028(0.095) & 0.171(0.069)\\
  &DT     & 0.408 & 0.249 & 0.248 & \underline{1.446} & \underline{1.053} & \underline{1.096} & 0.510(\textbf{-0.099}) & 0.300(\textbf{-0.050}) & 0.470(\textbf{-0.057})\\
  &SVM  & 0.650 & 0.148 & 0.208 & \underline{1.069} & \underline{1.142} & 0.959 & 0.777(\textbf{-0.076}) & 0.232(\textbf{-0.035}) & 0.434(\textbf{-0.033})\\
  &BN  & 0.498 & 0.528 & 0.459 & 0.958 & 0.882 & 0.967 & 0.378(0.056) & \underline{0.000}(\textbf{-0.422}) & \underline{0.001}(\textbf{-0.441}) \\    \hline
\multirow{6}*{Scen. 3}
    &J48   & 0.504 & 0.647 & 0.531 & 0.984 & \underline{1.007} & 0.990 & 0.884(\textbf{-0.019})& 0.614(\textbf{-0.002}) & 0.709(0.024) \\
    &LR    & 0.508 & 0.598 & 0.530 & 0.900 & \underline{1.002} & 0.963 & 0.784(0.024) & 0.601(\textbf{-0.024}) & 0.693(0.021)\\
    &NB   & 0.469 & 0.674 & 0.551 & 0.950 & \underline{1.030} & 0.951 & 0.886(0.003)& 0.465(\textbf{-0.049}) & 0.627(\textbf{-0.007})\\
    &DT   & 0.523 & 0.677 & 0.532 & 0.936 & 0.955 & \underline{1.012} & 0.277(0.059) & 0.131(\textbf{-0.085}) & 0.291(\textbf{-0.003})\\
    &SVM  & 0.518 & 0.623 & 0.539 & 0.916 & \underline{1.048} & \underline{1.002} & \underline{0.007}(0.075) & 0.322(\textbf{0.012}) & \underline{0.008}(0.050)\\
    &BN  & 0.480 & 0.661 & 0.530 & 0.924 & \underline{1.002} & 0.994 & 0.241(\textbf{-0.035}) & 0.879(\textbf{-0.007}) & 0.097(\textbf{-0.030}) \\   \hline
\end{tabular}
\end{table*}

\begin{table*}\small
\centering
\caption{The performance of the predictors built with \emph{mRMR} and \emph{TOP5} metrics and the comparison of the distributions of their results in \protect\\ terms of the Wilcoxon signed-rank test and Cliff's effect size ($d$): the underlined numbers and the negative numbers in bold represent the same meaning as in Table \ref{Simple1}.}\label{mRMR}
\begin{tabular}{c|c|c|c|c|c|c|c|c|c|c} \hline
 \multicolumn{2}{c|}{\multirow{2}*{ }}
 &\multicolumn{3}{c|}{Median \emph{mRMR} value} &\multicolumn{3}{c|}{Median \emph{TOP5}/\emph{mRMR} } &\multicolumn{3}{c}{\emph{mRMR} vs. \emph{TOP5} ($Sig.p <0.01$ ($d$))}
 \\
 \cline{3-11}
 \multicolumn{2}{c|}{}
  & \multicolumn{1}{c|}{Precision} &\multicolumn{1}{c|}{Recall} &\multicolumn{1}{c|}{F-measure}
  & \multicolumn{1}{c|}{Precision} &\multicolumn{1}{c|}{Recall} &\multicolumn{1}{c|}{F-measure}
  & \multicolumn{1}{c|}{Precision} &\multicolumn{1}{c|}{Recall} &\multicolumn{1}{c}{F-measure}
   \\ \hline
\multirow{6}*{Scen. 1}
    &J48  & 0.525 & 0.205 & 0.197 & 0.959 & \underline{1.881}  & \underline{1.150} & 0.546(\textbf{-0.052}) & 0.018(\textbf{-0.106}) & 0.030(\textbf{-0.123})\\
    &LR   & 0.634 & 0.280 & 0.253 & 0.955 & \underline{1.083} & 0.903 & 0.114(0.098) & 0.935(0.024) & 0.072(0.061) \\
    &NB   & 0.491 & 0.428 & 0.378 & \underline{1.035} & \underline{1.389}  & \underline{1.075} & 0.668(0.038) & \underline{0.000}(\textbf{-0.253}) & \underline{0.001}(\textbf{-0.186})\\
    &DT   & 0.289 & 0.245 & 0.203 & \underline{1.867} & \underline{1.296} & \underline{1.269} & 0.295(\textbf{-0.179}) & 0.064(\textbf{-0.089}) & 0.218(\textbf{-0.075}) \\
    &SVM  & 0.591 & 0.199 & 0.171 & 0.964 & \underline{1.412} & \underline{1.647} & 0.107(\textbf{-0.097}) & 0.167(\textbf{-0.071}) & 0.100(\textbf{-0.099})\\
    &BN  & 0.450 & 0.353 & 0.241 & \underline{1.078} & \underline{1.321} & \underline{1.847} & 0.136(\textbf{-0.134}) & 0.113(\textbf{-0.104}) & 0.117(\textbf{-0.116}) \\   \hline
\multirow{6}*{Scen. 2}
  &J48    & 0.394 & 0.176 & 0.203 & \underline{1.723}  & \underline{1.728}  & 0.981 & 0.073(\textbf{-0.226}) & 0.048(\textbf{-0.095}) & 0.394(0.076)\\
  &LR     & 0.690 & 0.250 & 0.243 & \underline{1.001}  & \underline{1.028} & 0.981 & 0.732(\textbf{-0.059}) & 0.758(0.049) & 0.241(0.069)\\
  &NB     & 0.522 & 0.377 & 0.376 & \underline{1.027}  & \underline{1.559} & \underline{1.174} & 0.841(0.049)& \underline{0.000}(\textbf{-0.226}) & \underline{0.002}(\textbf{-0.196})\\
  &DT     & 0.312 & 0.093 & 0.138 & \underline{1.889} & \underline{2.825} & \underline{1.971} & 0.159(\textbf{-0.255}) & 0.019(\textbf{-0.149}) & 0.039(\textbf{-0.175})\\
  &SVM  & 0.621 & 0.098 & 0.137 & \underline{1.119} & \underline{1.727} & \underline{1.455} & 0.230(\textbf{-0.130}) & 0.112(\textbf{-0.113}) & 0.108(\textbf{-0.128})\\
  &BN  & 0.472 & 0.266 & 0.312 & \underline{1.01} & \underline{1.755} & \underline{1.42} & 0.178(\textbf{-0.170}) & \underline{0.005}(\textbf{-0.224}) & 0.028(\textbf{-0.198}) \\    \hline
\multirow{6}*{Scen. 3}
    &J48   & 0.415 & 0.513 & 0.423 & \underline{1.195} & \underline{1.270} & \underline{1.242} & 0.301(\textbf{-0.085})& 0.099(\textbf{-0.191}) & 0.083(\textbf{-0.116}) \\
    &LR    & 0.403 & 0.552 & 0.456 & \underline{1.132} & \underline{1.086} & \underline{1.121} & 0.543(\textbf{-0.118}) & 0.153(\textbf{-0.094}) & 0.429(\textbf{-0.056})\\
    &NB   & 0.469 & 0.447 & 0.459 & 0.951 & \underline{1.550} & \underline{1.152} & 0.230(0.035)& \underline{0.000}(\textbf{-0.530}) & \underline{0.002}(\textbf{-0.215})\\
    &DT   & 0.471 & 0.551 & 0.492 & \underline{1.038} & \underline{1.174} & \underline{1.094} & 0.440(\textbf{-0.038}) & \underline{0.009}(\textbf{-0.161}) & 0.016(\textbf{-0.104})\\
    &SVM  & 0.482 & 0.525 & 0.461 & 0.984 & \underline{1.245} & \underline{1.172} & 0.475(0.017) & 0.207(\textbf{-0.137}) & 0.587(\textbf{-0.087})\\
    &BN  & 0.440 & 0.585 & 0.496 & \underline{1.006} & \underline{1.132} & \underline{1.062 }& 0.886(\textbf{0.028}) & 0.170(\textbf{-0.080}) & 0.361(\textbf{-0.076}) \\   \hline
\end{tabular}
\end{table*}

\subsection{RQ3: Which classifier is more likely to be the choice of defect prediction with a simplified metric set?}

Among all the predictors based on the six classifiers we studied, as shown in Figure \ref{Fig.2}, the Na\"{i}ve Bayes classifier provides the best median recall and F-measure in Scenarios 1 and 2. Although the precision presents a decreasing trend, it is rational to deem Na\"{i}ve Bayes as the most suitable classifier for WPDP in terms of \emph{Recall} or \emph{F-measure}. However, Logistic Regression or SVM is more likely to be the preferable choice for prediction models focusing on \emph{Precision}. In regard to CPDP, Decision Table appears to be the best classifier because of high recall, whereas Na\"{i}ve Bayes is another suitable classifier.

Furthermore, for the simplified metric sets obtained by other feature selection methods, Table \ref{MaxRel} and Table \ref{mRMR} indicates that the Na\"{i}ve Bayes classifier also tends to have greater measure values on the whole, and Bayesian Network shows comparable results in different scenarios. As we expected, Decision Table is also a preferable classifier for CPDP compared with Na\"{i}ve Bayes and Bayesian Network classifiers; SVM is preferred for WPDP with respect to \emph{Precision}.

\subsection{RQ4: Is there a minimum metric subset that facilitates the procedure for general defect prediction?}

\subsubsection{Minimizing the Top 5 metric subset}
In \emph{RQ2}, we validated that the \emph{TOP5} metric subset performs well according to the great representativeness and approximate comparability. However, there are still some strong correlations among the top five metrics. It is necessary to minimize the feature subset by eliminating the metrics that have a strong correlation with the others. According to the guideline shown in Table \ref{range}, a strong correlation between two metrics is identified as long as the correlation coefficient $r$ is greater than 0.6. Thus, $\varphi=0.6$ is selected as the threshold in the following experiment. Table \ref{Matrix} presents three correlation coefficient matrices in which four pairs of metrics have strong correlations, for example, the correlation between CBO and CE, and the correlations between RFC and LCOM, CE, and LOC. In particular, the correlation coefficient between RFC and LOC is greater than 0.9 in Scenario 1 and Scenario 2. Note that, we calculate the correlation coefficients between the top 5 metrics for the training data of each target release, and use their corresponding median values in our experiment.

\begin{table}\small
  \centering
  \caption{The correlation coefficient matrix $R^{k\times k}$, and the strong correlations are underlined.}\label{Matrix}
    \begin{tabular}{c|c|c|c|c|c}
      \hline
      Scenario1 & CBO & RFC & LCOM & CE & LOC \\ \hline
      CBO & 1 & 0.487 & 0.395 & \underline{0.622} & 0.379 \\
      RFC & - & 1 & \underline{0.616} & \underline{0.682} & \underline{0.909} \\
      LCOM & - & - & 1 & 0.375 & 0.49 \\
      CE & - & - & - & 1 & 0.587 \\
      LOC & - & - & - & - & 1 \\
      \hline
      Scenario2 & CBO & RFC & LCOM & CE & LOC \\ \hline
      CBO & 1 & 0.483 & 0.390 & \underline{0.617} & 0.375 \\
      RFC & - & 1 & \underline{0.611} & \underline{0.687} & \underline{0.910} \\
      LCOM & - & - & 1 & 0.373 & 0.487 \\
      CE & - & - & - & 1 & 0.592 \\
      LOC & - & - & - & - & 1 \\
      \hline
      Scenario3 & CBO & RFC & LCOM & CE & LOC \\ \hline
      CBO & 1 & 0.473 & 0.348 & \underline{0.640} & 0.352 \\
      RFC & - & 1 & \underline{0.601} & \underline{0.666} & \underline{0.879} \\
      LCOM & - & - & 1 & 0.337 & 0.423 \\
      CE & - & - & - & 1 & 0.544 \\
      LOC & - & - & - & - & 1 \\
      \hline
    \end{tabular}
\end{table}

\begin{figure}
  \centering
  \includegraphics[width=2.5in]{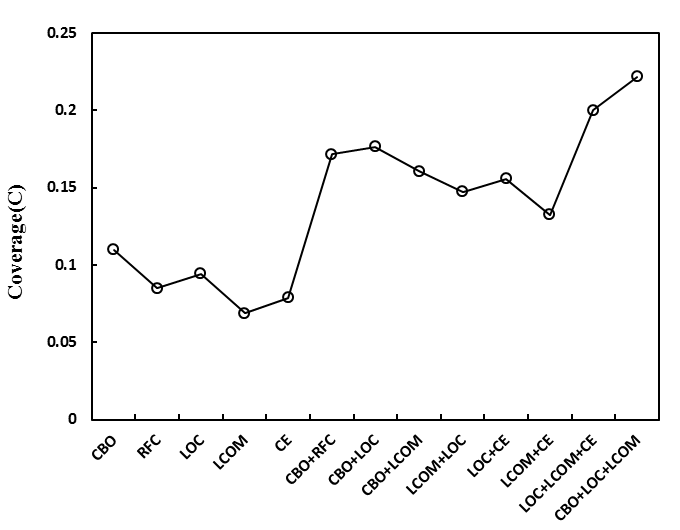}\\
  \caption{The \emph{Coverage} values of the remaining combinations. }\label{coverage}
\end{figure}

For the purpose of minimizing the simplified metric set, we eliminated any combinations that include those  metrics with strong correlations from the possible combinations ($C_{5}^{1}+C_{5}^{2}+C_{5}^{3}+C_{5}^{4}+C_{5}^{5}=31$ ), and the $C_{5}^{5}$ combination is excluded because it is identical to the \emph{TOP5} case. Considering the strong correlations, such as CBO vs. CE, RFC vs. LCOM, RFC vs. CE, and RFC vs. LOC, only thirteen of thirty combinations are remained when eliminating those combinations that contain at least one of the four pairs of metrics. Interestingly, most of the results calculated by the \emph{MaxRel} and \emph{mRMR} approaches are included in the 13 combinations after removing those metrics with strong correlations. At last, we calculate their \emph{Coverage} values to determine the suitable minimum metric subset. In Figure \ref{coverage}, the \emph{Coverage} values of those combinations that contain multiple metrics are obviously larger than that of one single metric, in particular, the combinations such as CBO+LOC, LOC+LCOM+CE and CBO+LOC+LCOM. To validate the existence of the minimum metric subset, the combination CBO+LOC+LCOM will be minutely explored in the following paragraphs.

\begin{figure*}
\centering
\includegraphics[width=6in,height=6in]{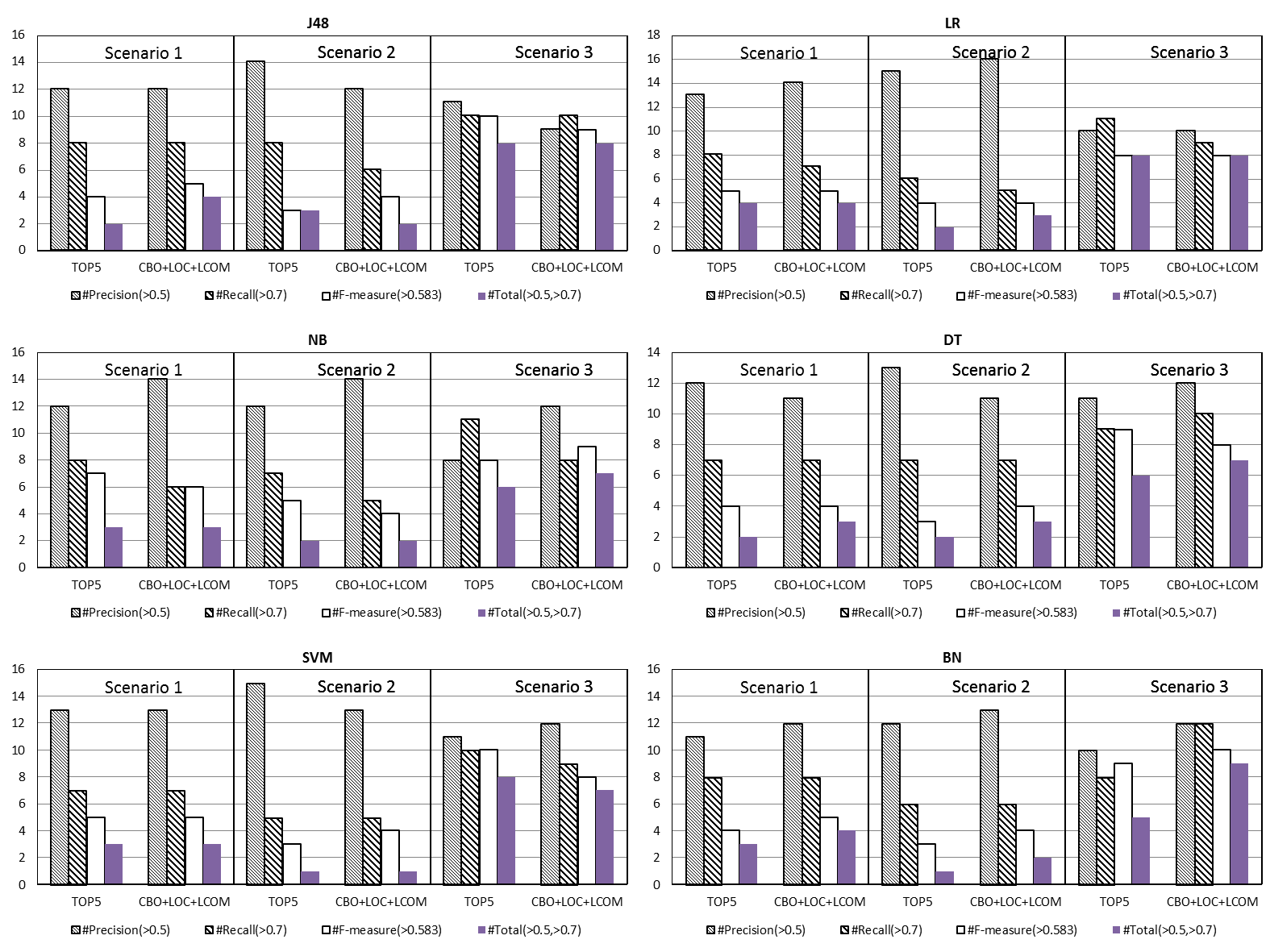}
\caption{The results of TOP5 and minimum metric subset under four conditions: $Precision> 0.5$, $Recall>0.7$,\\ \emph{F-measure}$>0.583$, and $Precision>0.5$ \& $Recall>0.7$. The Y-axis is the number of results under the given condition.}
\label{compare}
\end{figure*}

\subsubsection{Prediction results of the predictor based on the minimum metric subset}

\begin{table*}\small
\centering
\caption{The performance of the predictors built with \emph{TOP5} and \emph{CBO+LOC+LCOM} (abbreviated as \emph{CLL}) metrics and the comparison of the distributions of their results in terms of the Wilcoxon signed-rank test and Cliff's effect size ($d$): the underlined numbers in the Median \emph{CLL/TOP5} column represent that the result of \emph{CLL} is better, and those in the \emph{TOP5} vs. \emph{CLL} column represent that one can reject the \emph{null hypothesis}; the negative numbers in bold represent the result of \emph{CLL} is better.}\label{CLL}
\begin{tabular}{c|c|c|c|c|c|c|c|c|c|c} \hline
 \multicolumn{2}{c|}{\multirow{2}*{ }}
 &\multicolumn{3}{c|}{ Median \emph{CLL} value} &\multicolumn{3}{c|}{Median \emph{CLL/TOP5} } &\multicolumn{3}{c}{\emph{TOP5} vs. \emph{CLL} ($Sig.p <0.01$ ($d$))}
 \\
 \cline{3-11}
 \multicolumn{2}{c|}{}
  & \multicolumn{1}{c|}{Precision} &\multicolumn{1}{c|}{Recall} &\multicolumn{1}{c|}{F-measure}
  & \multicolumn{1}{c|}{Precision} &\multicolumn{1}{c|}{Recall} &\multicolumn{1}{c|}{F-measure}
  & \multicolumn{1}{c|}{Precision} &\multicolumn{1}{c|}{Recall} &\multicolumn{1}{c}{F-measure}
   \\ \hline
\multirow{6}*{Scen. 1}
    &J48  & 0.524 & 0.342 & 0.264 & \underline{1.041} & 0.888 & \underline{1.164} & 0.709(0.007) & 0.351(0.045) & 0.903(0.005)\\
    &LR   & 0.654 & 0.209 & 0.189 & \underline{1.08} & 0.687 & 0.830 & 1.000(\textbf{-0.052}) & 0.023(0.069) & 0.100(0.063) \\
    &NB   & 0.597 & 0.430 & 0.338 & \underline{1.173} & 0.732  & 0.832 & 0.189(\textbf{-0.063}) & \underline{0.000}(0.189) & \underline{0.001}(0.181)\\
    &DT   & 0.413 & 0.342 & 0.285 & 0.766 & \underline{1.076} & \underline{1.108} & 0.388(0.149) & 0.701(0.019) & 0.650(0.032) \\
    &SVM  & 0.607 & 0.244 & 0.202 & \underline{1.066} & 0.869 & 0.718 & 0.695(\textbf{-0.043}) & 0.151(0.017) & 0.289(0.017) \\
    &BN  & 0.500 & 0.377 & 0.343 & \underline{1.03} & 0.810 & 0.771 & 0.877(0.031) & 0.177(0.057) & 0.163(0.085) \\  \hline
\multirow{6}*{Scen. 2}
  &J48    & 0.532 & 0.225 & 0.223 & 0.784 & 0.738 & \underline{1.118} & 0.355(0.158) & 0.968(0.043) & 0.664(0.031)\\
  &LR     &0.700 & 0.184 & 0.195 & \underline{1.015}  & 0.715 & 0.817 & 0.958(0.068) & 0.013(0.078) & 0.082(0.073)\\
  &NB     &0.560 & 0.420 & 0.376 & \underline{1.04}  & 0.715 & 0.851 & 0.028(\textbf{-0.104})& \underline{0.000}(0.241)& \underline{0.000}(0.205)\\
  &DT    &0.351 & 0.319 & 0.257 & 0.595 & \underline{1.216} & 0.946 & 0.239(0.189) & 0.937(0.030) & 0.875(0.043)\\
  &SVM  & 0.646 & 0.122 & 0.182 & 0.930 & 0.719 & 0.914 & 0.332(0.115) & 0.446(0.052) & 0.411(0.073) \\
  &BN  & 0.552 & 0.372 & 0.321 & \underline{1.158} & 0.798 & 0.725 & 0.794(0.028) & 0.017(0.203) & 0.010(0.191) \\  \hline
\multirow{6}*{Scen. 3}
    &J48  & 0.472 & 0.630 & 0.532 & 0.952 & 0.968 & \underline{1.011} & 0.503(0.069)& 0.648(0.043) & 0.627(0.009) \\
    &LR   & 0.453 & 0.573 & 0.500 & 0.991 & 0.957 & 0.979 & 0.584(0.080) & 0.259(0.030) & 0.201(0.024)\\
    &NB   & 0.501 & 0.622 & 0.535 & \underline{1.124} & 0.897 & \underline{1.020} & 0.230(\textbf{-0.052})& 0.223(0.104) & 0.886(0.017)\\
    &DT   & 0.494 & 0.674 & 0.538 & \underline{1.009} & \underline{1.042} & 0.999 & 0.715(0.002) & 0.885(\textbf{-0.050}) & 0.664(\textbf{-0.012})\\
    &SVM  & 0.484 & 0.595 & 0.531 & \underline{1.021} & 0.911 & 0.982 & 0.668(0.000) & 0.256(0.069) & 0.253(0.038)\\
    &BN  & 0.493 & 0.710 & 0.532 & \underline{1.114} & \underline{1.072} & \underline{1.009} & 0.429(\textbf{-0.016}) & 0.230(\textbf{-0.118}) & 0.543(\textbf{-0.043}) \\ \hline
\end{tabular}
\end{table*}

First, we have to determine the corresponding thresholds of \emph{Recall}, \emph{Precision} and \emph{F-measure} that are to be adopted to evaluate the minimum subset. Like the literature \cite{He:An}, in our study, the thresholds 0.5 and 0.7 were selected for \emph{Precision} and \emph{Recall} respectively. As a weighted average of \emph{Precision} and \emph{Recall}, a value of 0.583 is used for \emph{F-measure}. Thus, we compared the results for different combinations with the six classifiers under four types of evaluation conditions based on the given thresholds (see Figure \ref{compare}). $\#Precision, \#Recall$, \emph{\#F-measure} and $\#Total$ indicate the number of results for a combination that meet the given threshold of their respective evaluation condition.

For Scenario 1, compared with the \emph{TOP5} metric subset, there is a non-decreasing trend in the number of results for CBO+LOC+LCOM under the given condition of $Precision$ except the Decision Table case. Four out of six cases still have equal $\#Recall$, though minimizing  metric set causes a decrease in $\#Recall$ for the cases of Logistic Regression and Na\"{\i}ve Bayes. \emph{\#F-measure} exhibits a slight upward trend except the Na\"{\i}ve Bayes case, especially for the cases of J48 and Bayesian Network. Moreover, the improvement is more optimistic when considering the precision greater than 0.5 and the recall greater than 0.7 together, indicated by an increase in the number of the results that meet both of the conditions. Generally, the results for the minimum metric subset are approximately equal to the $TOP5$ metric subset.

For Scenario 2, a similar phenomenon of the results under the given condition of \emph{F-measure} when using all the historical data is shown in Figure \ref{compare}, but the fluctuation of the precision presents a different result compared with using the nearest historical data. At the same time, the results confirm our previous finding obtained in \emph{RQ1}: the quantities of training data do not remarkably affect the prediction results; on the contrary, a slightly worse trend under the given condition of $Recall$ is presented in this scenario. Additionally, we also find that the minimum metric subset (i.e., CBO+LOC+LCOM) could be selected as an alternative choice because of the comparative results under the given conditions of \emph{F-measure} and $Total$.

Interestingly, for Scenario 3, some results observed in the other two scenarios are confirmed again. For example, compared with the $TOP5$ metric subset, there is a non-decreasing trend in the number of results under the given condition of $Precision$ except one case. In particular, the results under the given condition of $Total$ still maintain competitive for CPDP. This scenario also presents several significant differences in terms of the number of results under the given conditions. For example, compared with WPDP, there is an obvious increase in $\#Recall$, \emph{\#F-measure} and $\#Total$ for CPDP. However, in regard to $\#Precision$, Scenario 3 exhibits a clear downward trend in comparison with Scenario 2, especially for the cases of J48 and Logistic Regression. The findings indicate that the minimum metric subset is also appropriate for CPDP with little loss of precision.

\begin{figure*} \small.
\centering
\includegraphics[width=5.5in,height=6in]{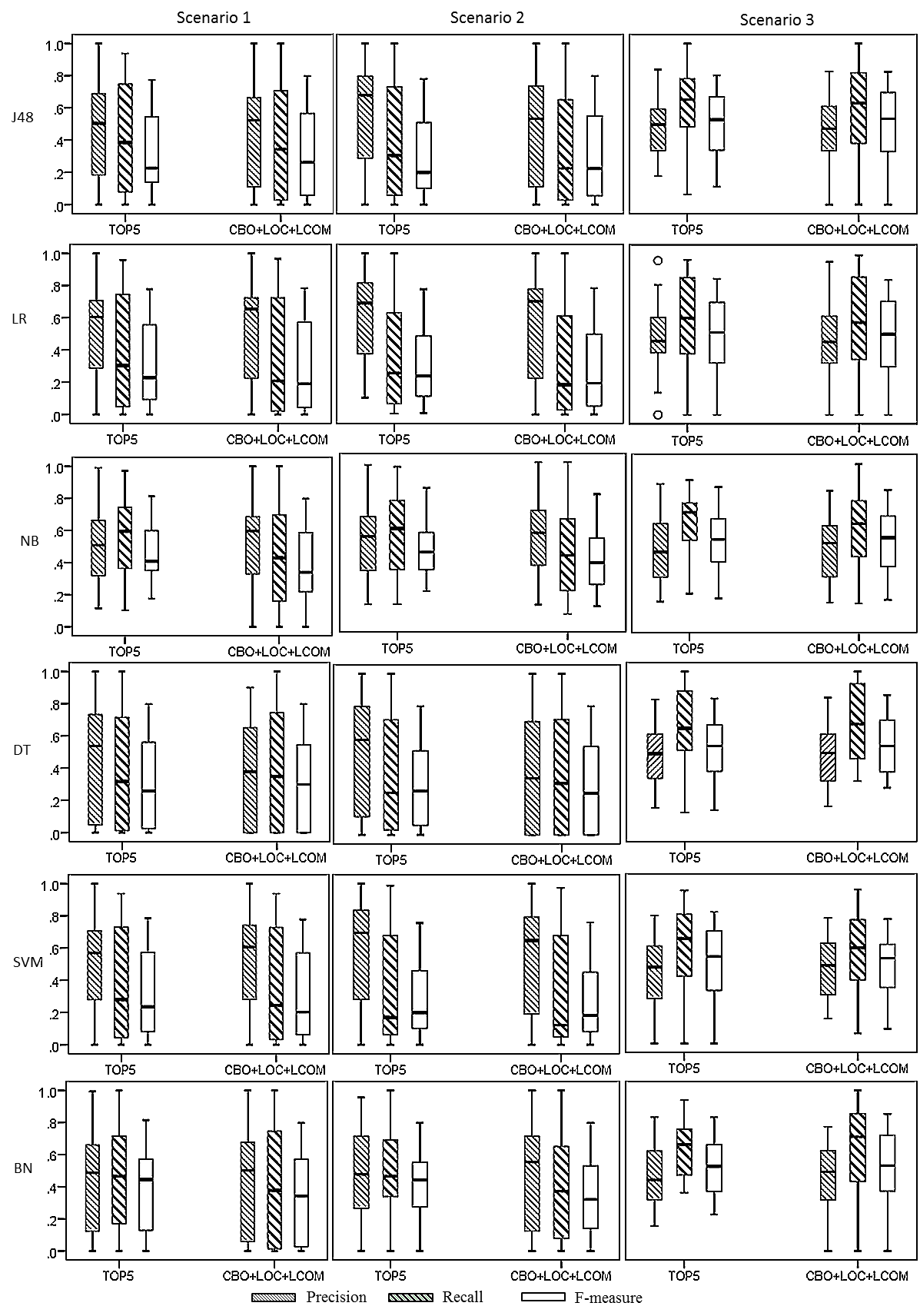}
\caption{The standardized boxplots of the performance achieved by the predictors built with TOP5 and minimum metric subset based on six classifiers in different scenarios, respectively. From the bottom to the top of a standard box plot: minimum, first quartile, median, third quartile, and maximum. Any data not included between the whiskers is plotted as a small circle.}
\label{compare2}
\end{figure*}

In addition, Figure \ref{compare2} presents the similar prediction results achieved by the predictors built with $TOP5$ and the minimum metric subset CBO+LOC+LCOM based on six typical classifiers in different scenarios, respectively. As another evidence, Table \ref{CLL} further shows a comparison between them in terms of the Wilcoxon signed-ranked test and Cliff's effect size. The results suggest that we cannot reject the \emph{null hypothesis} on the whole, implying that there is no statistically significant difference between them. However, we have to admit that, for WPDP based on the Na\"{\i}ve Bayes classifier, the recall and F-measure of the minimum metric subset do not appear to be as good as the $TOP5$ metric subset, whereas the former outperforms the latter in terms of the precision.

CBO+LOC+LCOM is, by far, empirically validated to be a basic metric set for defect prediction both within a project and across projects. An explanation of our finding can be underpinned  by the facts that (1) the Pareto principle of defect distribution, (i.e., a small number of modules account for a large proportion of the defects found \cite{Fenton: Quantitative,Ostrand: The,Andersson: A}); (2) larger modules tend to have more defects (i.e., a strong positive correlation exists between LOC and defects \cite{Zhang:An}); (3) as the representative complexity metrics, both CBO and LCOM are important indicators for fault-prone classes in terms of software coupling and cohesion, which have been commonly used to defect-proneness prediction \cite{Subramanyam: Empirical,Gyimothy: Empirical,Ma: A}.

\subsubsection{Stability of the minimum metric subset}

To understand the stability of prediction results of the predictor built with the minimum metric subset, we performed ANalysis Of VAriance (ANOVA) \cite{Berenson:Intermediate} to statistically validate the robustness (i.e., consistency) of such a predictor. An n-way ANOVA can be used to determine whether a significant impact on the mean in a set of data is developed by multiple factors. In this study, we used one-way ANOVA with a single factor, which is the choice among the six classifiers. For this ANOVA, we first calculated the \emph{Consistency} value of each classifier with the $TOP5$ metric subset, CBO+LOC (see \emph{RQ4} in the section of discussion) and the minimum metric subset according to Equation (\ref{Eq.3}), which was defined in Section 4.5. Then, ANOVA was used to examine the hypothesis that the \emph{Consistency} values of  simplified metric subsets for all classifiers are equal against the alternative hypothesis that at least one mean is different. The test of statistical significance in this section utilized a significance level of $p<0.05$, and the whole test was implemented in IBM SPSS Statistics.

The ANOVA results are presented in Table \ref{anova}. The $p$ values are greater than 0.1, which indicates that there is no significant difference of the average \emph{Consistency} values among these six classifiers. That is, the minimum metric subset is relatively stable and can be independent of classifiers. Note that, the good stability of the minimum metric subset dose not contradict the finding that the Na\"{\i}ve Bayes classifier is more likely to be the choice of defect prediction with a simplified metric subset obtained in \emph{RQ3}.

\begin{table*}\small
\centering
\caption{ANalysis Of VAriance for Consistency results among different classifiers.}\label{anova}
\begin{tabular}{c|c|c|c|c|c|c}
  \hline
  Scenario & Combination & Sum of Squares & d.f.& Mean Square & F & \emph{p}-value \\ \hline
\multirow{3}*{1}
    & TOP5    & 1.373 & 5 & 0.275 & 0.445 & 0.816  \\
    & CBO+LOC & 0.145 & 5 & 0.029 & 0.038 & 0.999 \\
    & CBO+LOC+LCOM & 0.309 & 5 & 0.062 & 0.091 & 0.994 \\
 \hline
\multirow{3}*{2}
    & TOP5    & 2.160 & 5 & 0.432 & 0.690 & 0.632  \\
    & CBO+LOC & 0.198 & 5 & 0.405 & 0.054 & 0.998 \\
    & CBO+LOC+LCOM & 0.531 & 5 & 0.106 & 0.167 & 0.974 \\
 \hline
\multirow{3}*{3}
    & TOP5    & 0.082 & 5 & 0.016 & 0.164 & 0.975  \\
    & CBO+LOC & 0.261 & 5 & 0.052 & 0.414 & 0.839 \\
    & CBO+LOC+LCOM & 0.756 & 5 & 0.151 & 0.1.601 & 0.164 \\
 \hline
\end{tabular}
\end{table*}

\subsection{A Summary of the Results}

In summary, the goal of this study is to investigate a simplified approach that can make a trade-off between generality, performance, and complexity. The primary issues that we focus on are (1) how to select training data sets for defect prediction, (2) how to determine the suitable simplified feature subset for defect prediction, and (3) whether or not simple learning algorithm tends to perform well in this context. Our analysis empirically validates the idea that there are certain guidelines available for reference to answer the proposed research questions. We particularly emphasize that our analysis supports the hypothesis that a predictor built with a simplified metric set, in different scenarios, is feasible and practical to predict defect-prone classes.

Table \ref{summary1} summarizes the results of our study. WPDP models generally capture high precision, whereas CPDP models always achieve higher recall and F-measure. The difference is significantly discriminated by the F-measure. For example, the median F-measure is either high or medium in CPDP, while the median values of this measure do not exceed the middle level in WPDP. In addition, in WPDP, using a simple classifier (Na\"{\i}ve Bayes) can improve the recall and maintain an appropriate level of precision. A simple classifier (e.g., Logistic Regression and Na\"{\i}ve Bayes) also performs well in CPDP with respect to the overall performance. In other words, in WPDP, Na\"{\i}ve Bayes provides high recall and Decision Table maintains stable results for different metric sets. In CPDP, Na\"{\i}ve Bayes and Bayesian Network are relatively stable for different metric sets, and Decision Table and SVM become suitable as selections for high F-measure values. Note that, the ``Times" column represents the number of occurrences of each combination of the three measures in Figure  \ref{Fig.2}. For instance, in Scenario 1, the frequency of the first combination (H, H, M) is 2. In other words, there are two prediction results that meet the specific performance requirement in this scenario. This is also the reason why there are several rows per scenario. The ``Predictors" column is used to characterize the condition which classifiers and types of metric sets are available for a specific performance requirement.

Considering the advantages of simple classifiers as mentioned in $RQ 3$, Table \ref{summary2} further summarizes a guideline for choosing the suitable metric sets to facilitate defect prediction with them such as Na\"{\i}ve Bayes. Software engineers have three choices to determine which metric subsets are suitable for implementing the specific requirements. For instance, if only appropriate precision (e.g., $Precision$ is around 0.5) is required for WPDP, he/she can select the metrics determined by the default feature selection technique in Weka. If both appropriate precision and high recall are required at the same time, \emph{TOPK} is preferentially recommended. Furthermore, if higher precision is required under the above conditions, the minimum metric subset (e.g., CBO+LOC +LCOM) is suggested as the best choice for engineers. Additionally, for CPDP, one can select the \emph{FILTER} metric subset determined by Weka when only high recall is required. If higher recall and high F-measure are required together, \emph{TOPK} is preferentially recommended. Finally, the minimum metric subset will be recommended if appropriate precision or high F-measure is required.

\begin{table*}\small
\newcommand{\tabincell}[2]{\begin{tabular}{@{}#1@{}}#2\end{tabular}}
\caption{A summary of the selection of training data, classifiers and software metric sets. \underline{H}igh($median>0.5$), \protect \\ \underline{M}edium($0.5\geq median>0.3$), and \underline{L}ow($median\leq 0.3$) are used to represent the requirement of performance, \protect\\ and the initial A, F, and T represent the corresponding metric set \emph{ALL}, \emph{FILTER}, and \emph{TOP5}, respectively.}\label{summary1}
\centering
\begin{tabular}{c|c|c|c|c|c|c|c|c|c|c|c}
  \hline
  \multicolumn{1}{c|}{\multirow{2}*{Scen.}} & \multicolumn{1}{c|}{\multirow{2}*{Training data}}
   & \multicolumn{4}{c|}{Performance Measures} & \multicolumn{6}{c}{Predictors} \\
\cline{3-12}
\multicolumn{1}{c|}{} & \multicolumn{1}{c|}{}
    & Precision & Recall & F-measure & Times & J48 & LR & NB & DT & SVM & BN \\
\hline
  \multicolumn{1}{c|}{\multirow{6}*{1} }& \multicolumn{1}{c|}{\multirow{6}*{\tabincell{c}{From the latest \\ release}}}
    & H & H   & M & 2 &   &   & F/T &  &  & \\
  &  & H & M & M & 1 & A & & & & &  \\
  &  & H & M & L & 6 & T & A/T &   & A/F/T &       & \\
  &  & H & L & L & 5 & F & F   &   &       & A/F/T &  \\
  &  & M & H & M & 1 &   &     & A &       &       & \\
  &  & M & M & M & 3 &   &     &   &       &       & A/F/T \\
\hline
  \multicolumn{1}{c|}{\multirow{6}*{2}} & \multicolumn{1}{c|}{\multirow{6}*{\tabincell{c}{From the existing \\ releases}}}
    & H & H & M & 2  &   &   & A/T & & & \\
  &  & H & M & L & 1  & T &   &     & & &  \\
  &  & H & L & M & 1  & A &   &     & & & \\
  &  & H & L & L & 10 & F & A/F/T &  & A/F/T & A/F/T & \\
  &  & M & H & M & 2  &   &   & F &  & & A  \\
  &  & M & M & M & 2  &   &   &   &  & & F/T  \\
\hline
  \multicolumn{1}{c|}{\multirow{3}*{3}} & \multicolumn{1}{c|}{\multirow{3}*{\tabincell{c}{From other \\ projects}}}
  & H & H & H & 4 &   &   &   & A/F & A/F & \\
  &  & M & H & H &10 & T & T & A/F/T & T & T & A/F/T \\
  &  & M & H & M & 4 & A/F & A/F &   &  & &  \\
  \hline
  \multicolumn{12}{c}{\tabincell{c}{$\ast$ The Times column represents the number of occurrences of the corresponding combination of the three measures in  Figure \ref{Fig.2}.}} \\
\end{tabular}
\end{table*}

\begin{table*}\small
\newcommand{\tabincell}[2]{\begin{tabular}{@{}#1@{}}#2\end{tabular}}
  \centering
  \caption{A guideline for choosing the suitable metric subset in different scenarios (ref. Figure \ref{Fig.2} and Figure \ref{compare2}).}\label{summary2}
\begin{tabular}{c|c|c|c|c|c|c|c}
  \hline
  \multicolumn{1}{c|}{\multirow{2}*{Scen.}} & \multicolumn{1}{c|}{\multirow{2}*{Training data}}
   & \multicolumn{3}{c|}{\multirow{2}*{Performance Requirements}} & \multicolumn{1}{c|}{\multirow{2}*{FILTER}}
   & \multicolumn{1}{c|}{\multirow{2}*{TOP K}} & \multicolumn{1}{c}{\multirow{2}*{Minimum}} \\
  \multicolumn{1}{c|}{} & \multicolumn{1}{c|}{} & \multicolumn{1}{c}{} & \multicolumn{1}{c}{}
  &  \multicolumn{1}{c|}{} &  \multicolumn{1}{c|}{} & \multicolumn{1}{c|}{}  & \multicolumn{1}{c}{}  \\
\hline
  \multicolumn{1}{c|}{\multirow{3}*{} } & \multicolumn{1}{c|}{\multirow{3}*{} }& \multicolumn{3}{c|}{\tabincell{c}{Appropriate precision and high recall \\are required
(e.g., (0.509, 0.594)). } } & $\diagdown$ & \tabincell{c}{support \\ (e.g., TOP5) }& $\diagdown$ \\
\cline{3-8}
  \multicolumn{1}{c|}{1} & \multicolumn{1}{c|}{\tabincell{c}{From the latest \\ release}} & \multicolumn{3}{c|}{\tabincell{c}{ High precision is required \\ (e.g., 0.535).}} & support & $\diagdown$ & $\diagdown$ \\
\cline{3-8}
  \multicolumn{1}{c|}{} & \multicolumn{1}{c|}{} & \multicolumn{3}{c|}{\tabincell{c}{Higher precision is required \\ (e.g., 0.597).}} & $\diagdown$ & $\diagdown$ & \tabincell{c}{support \\ (e.g., CBO+LOC+LCOM)}\\
\hline
 \multicolumn{1}{c|}{\multirow{3}*{} } & \multicolumn{1}{c|}{\multirow{3}*{} } & \multicolumn{3}{c|}{\tabincell{c}{Appropriate precision is required \\ (e.g., 0.494).  } } & weakly support& $\diagdown$ & $\diagdown$ \\
\cline{3-8}
  \multicolumn{1}{c|}{2} & \multicolumn{1}{c|}{\tabincell{c}{From historical \\ releases}} & \multicolumn{3}{c|}{\tabincell{c}{High precision and recall are required \\ (e.g., (0.536, 0.587)). } } & $\diagdown$ & \tabincell{c}{support \\ (e.g., TOP5) } & $\diagdown$ \\
\cline{3-8}
  \multicolumn{1}{c|}{} & \multicolumn{1}{c|}{} & \multicolumn{3}{c|}{\tabincell{c}{Higher precision is required \\ (e.g., 0.560).}} & $\diagdown$ & $\diagdown$ & \tabincell{c}{support \\(e.g., CBO+LOC+LCOM) }\\
\hline
 \multicolumn{1}{c|}{\multirow{3}*{} } & \multicolumn{1}{c|}{\multirow{3}*{} }& \multicolumn{3}{c|}{\tabincell{c}{Appropriate precision (e.g., 0.501)\\ or high F-measure is required (e.g., 0.535).}} & $\diagdown$ & $\diagdown$  & \tabincell{c}{support \\(e.g., CBO+LOC+LCOM)} \\
\cline{3-8}
  \multicolumn{1}{c|}{3} & \multicolumn{1}{c|}{\tabincell{c}{From other \\ projects}} & \multicolumn{3}{c|}{ \tabincell{c}{High recall is  required \\(e.g., 0.66).} } & support  &  $\diagdown$ & $\diagdown$ \\
\cline{3-8}
   \multicolumn{1}{c|}{} & \multicolumn{1}{c|}{} &  \multicolumn{3}{c|}{\tabincell{c}{Higher recall and high F-measure are \\required (e.g., (0.694, 0.525)).}} & $\diagdown$ & \tabincell{c}{support \\ (e.g., TOP5) } & $\diagdown$ \\
\hline
\end{tabular}
\end{table*}

\begin{table*}\small
\centering
\caption{The performance of the predictors built with \emph{CBO+LOC} (abbreviated as \emph{CL}) and \emph{CBO+LOC+LCOM} (abbreviated as \emph{CLL}) metrics and the comparison of the distributions of their results in terms of the Wilcoxon signed-rank test and Cliff's effect size ($d$): the underlined numbers and the negative numbers in bold represent that the result of \emph{CL} is better.}\label{LOCBO}
\begin{tabular}{c|c|c|c|c|c|c|c|c|c|c} \hline
 \multicolumn{2}{c|}{\multirow{2}*{ }}
 &\multicolumn{3}{c|}{ Median \emph{CL} value} &\multicolumn{3}{c|}{Median \emph{CL/CLL} } &\multicolumn{3}{c}{\emph{CLL} vs. \emph{CL} ($Sig.p <0.01$ ($d$))}
 \\
 \cline{3-11}
 \multicolumn{2}{c|}{}
  & \multicolumn{1}{c|}{Precision} &\multicolumn{1}{c|}{Recall} &\multicolumn{1}{c|}{F-measure}
  & \multicolumn{1}{c|}{Precision} &\multicolumn{1}{c|}{Recall} &\multicolumn{1}{c|}{F-measure}
  & \multicolumn{1}{c|}{Precision} &\multicolumn{1}{c|}{Recall} &\multicolumn{1}{c}{F-measure}
   \\ \hline
\multirow{6}*{Scen. 1}
    &J48  & 0.424 & 0.365 & 0.285 & 0.808 & \underline{1.068} & \underline{1.079} & 0.158(0.144) & 0.551(0.043) & 0.551(0.067)\\
    &LR   & 0.641 & 0.169 & 0.163 & 0.980 & 0.812 & 0.863 & 0.603(\textbf{-0.005}) & 0.586(0.010) & 0.349(0.028) \\
    &NB   & 0.618 & 0.315 & 0.247 & \underline{1.036} & 0.733  & 0.731 & 0.401(0.005) & 0.295(0.083) & 0.068(0.153)\\
    &DT   & 0.319 & 0.383 & 0.292 & 0.774 & \underline{1.120} & \underline{1.024} & 0.398(0.072) & 0.116(\textbf{-0.049}) & 0.310(\textbf{-0.009}) \\
    &SVM  & 0.640 & 0.184 & 0.208 & \underline{1.053} & 0.753 & \underline{1.029} & 0.526(0.083) & 0.526(0.080) & 0.149(0.094) \\
    &BN  & 0.319 & 0.342 & 0.292 & 639 & 0.906 & 0.851 & 0.477(0.128) & 0.534(0.036) & 0.286(0.071) \\  \hline
\multirow{6}*{Scen. 2}
  &J48    & 0.431 & 0.248 & 0.262 & 0.809 & \underline{1.105} & \underline{1.174} & 0.279(0.109) & 0.701(\textbf{-0.021}) & 0.600(\textbf{-0.014})\\
  &LR     &0.690 & 0.172 & 0.213 & 0.986  & 0.936 & \underline{1.093} & 0.542(\textbf{-0.023}) & 0.177(0.040) & 0.177(0.030)\\
  &NB     &0.618 & 0.231 & 0.238 & \underline{1.104}  & 0.551 & 0.633 & 0.527(\textbf{-0.021})& 0.030(0.205)& 0.010(0.273)\\
  &DT    &0.058 & 0.112 & 0.102 & 0.165 & 0.350 & 0.397 & 0.169(0.226) & 0.066(0.102) & 0.022(0.142)\\
  &SVM  & 0.552 & 0.116 & 0.147 & 0.855 & 0.956 & 0.809 & 0.876(0.082) & 0.744(0.095) & 0.543(0.083) \\
  &BN  & 0.215 & 0.274 & 0.223 & 0.390 & 0.738 & 0.693 & 0.124(0.262) & 0.191(0.121) & 0.078(0.191) \\  \hline
\multirow{6}*{Scen. 3}
    &J48  & 0.461 & 0.624 & 0.538 & 0.977 & 0.990 & \underline{1.011} & 0.795(0.009)& 0.352(\textbf{-0.012}) & 0.831(0.002) \\
    &LR   & 0.423 & 0.597 & 0.517 & 0.935 & \underline{1.043} & \underline{1.033} & 0.543(0.005) & 0.351(\textbf{-0.031}) & 0.768(\textbf{-0.019})\\
    &NB   & 0.452 & 0.599 & 0.530 & 0.902 & 0.963 & 0.992 & 0.265(\textbf{-0.045})& 0.966(\textbf{-0.014}) & 0.587(0.000)\\
    &DT   & 0.494 & 0.666 & 0.540 & \underline{1.000} & 0.988 & \underline{1.004} & 0.091(\textbf{-0.057}) & 0.075(0.050) & 0.182(0.049)\\
    &SVM  & 0.482 & 0.649 & 0.549 & 0.995 & \underline{1.090} & \underline{1.035} & 0.241(\textbf{-0.035}) & 0.181(\textbf{-0.095}) & 0.052(\textbf{-0.080})\\
    &BN  & 0.462 & 0.674 & 0.534 & 0.936 & 0.949 & \underline{1.004} & 0.041(\textbf{-0.047}) & 0.041(0.066) & 0.100(0.019) \\ \hline
\end{tabular}
\end{table*}

\section{Discussion}
\emph{RQ1:} Our experimental results described in the previous section validate that CPDP is feasible for a project with limited data sets, and it even performs better than WPDP in terms of \emph{Recall} and \emph{F-measure}. For CPDP, we must state that, in this paper, the combinations of the most suitable training data from other projects are selected based on the approach proposed in \cite{He:An} in an exhaustive way. All the combinations of data sets used to train prediction models consist of no more than three releases from other projects. One reasonable explanation is that almost all projects in question have no more than four releases. For a more detailed description of the method that uses the most suitable training data from other projects, please refer to \cite{He:An}.

Defect prediction performs well as long as there is a sufficient amount of data available to train any models \cite{Zimmermann:Cross}, whereas it does not mean that more data must lead to more precise prediction models. We find that there is no observable improvement when increasing the number of training data sets in WPDP. Therefore, the quality of the training data is more important than the amount during defect prediction.

\emph{RQ2:} To the best of our knowledge, there is no widely accepted standard for judging the desired generalization and accuracy of different prediction models. We used statistical methods to examine whether there is a significant difference between the predictor built with a simplified metric set and a benchmark predictor in terms of evaluation measures. Although this treatment may be subjective to determine what is acceptable, the thresholds of median ratio 0.9 and nonparametric test (the Wilcoxon signed-rank test and Cliff's effect size) are meaningful and reliable in practice.

Either the method without using feature selection techniques or the approach based on \emph{filters} is commonly used to build defect predictors. Considering the simple pursuit of prediction precision, these methods are not the best reference models. However, their primary merit is independent of specific learning algorithms, suggesting that they are repeatable, versatile, and easy to use. Thus, we argue that they are suitable references for our empirical study from an overall viewpoint of performance. In addition, we used the top five metrics to build the third predictor (\emph{TOPK}) with regard to the three factors, and the comparison of different lengths \emph{K} for \emph{Coverage} values is presented in Figure \ref{occurrence}. In this figure, we can observe that the \emph{Coverage} value reaches a peak when $K=5$. A proper size is very important to simplify the metric set for defect prediction.

\emph{RQ3:} We find that simple classifiers (e.g., Na\"{\i}ve Bayes) tend to perform better on the whole when using a simplified metric set for defect prediction in all three scenarios. The result is completely consistent with the conclusions proposed in the literature \cite{Hall:A, Catal:Software}. Specifically, Na\"{\i}ve Bayes is a robust machine learning algorithm for supervised software defect prediction problems in both WPDP and CPDP. However, we have to admit that it is not suitable for some specific performance requirements. For example, for WPDP, SVM outperforms Na\"{\i}ve Bayes with respect to \emph{Precision}.

\emph{RQ4:}  LOC, CBO, and LCOM are considered to be suitable components of the minimum metric subset, which is largely consistent with the results of several prior studies. For LOC, Zhang \cite{Zhang:An} has performed a detailed investigation of the relationship between LOC and defects. Their study confirmed that a simple static code attribute, such as LOC, could be a useful predictor of software quality. CBO measures the degree of coupling between object classes and LCOM measures how well the methods of a class are related to each other. High cohesion and low coupling are two basic principles of software engineering. These three metrics may be appropriate features for defect prediction in our context.

Interestingly, the metric subset CBO+LOC, as a subset of CBO+LOC+LCOM, not only has the highest \emph{Coverage} value among the combinations with two metrics, but also can achieve a similar result that has no statistically significant difference compared with CBO+LOC+LCOM in terms of the Wilcoxon signed-rank test (see Table \ref{LOCBO}). According to Table \ref{LOCBO}, CBO+LOC seems to perform well in Scenario 3, but the effect is mediocre in the other two scenarios. On the other hand,  Table \ref{anova} also shows that the metric subset CBO+LOC is relatively stable for different classifiers. Therefore, the metric subset CBO+LOC could be an alternative choice of the minimum metric subset for CPDP from a practical point of view, even though the combination CBO+LOC+LCOM has been validated as the minimum metric subset from a theoretical point of view.

As an alternative to handling several metrics, a simplified approach to predicting defectiveness, which is practical and easy, could determine the subset of metrics that are cardinal and determine the correct organization \cite{Okutan:Software}. This approach was validated using the Wilcoxon signed-rank test and Cliff's effect size ($d$) in both WPDP and CPDP (see Table \ref{Simple1}, Table \ref{Simple2}, and Table \ref{CLL}). The implications of using a simplified metric set for defect prediction are effective reduction of the cost of data acquisition and processing by sacrificing a little performance. According to our study, the results indicate that the advantages of this approach outweigh the disadvantages.

\section{Threats to Validity}
In this study, we obtained several significant results to answer the four research questions proposed in Section 3.2. However, potential threats to the validity of our work still remain.

Threats to \emph{construct validity} concern the relationship between theory and observation. These threats are primarily related to the static code metrics we used. All the data sets were collected by Jureczko and Madeyski \cite{Jureczko:Towards}, and Jureczko and Spinellis \cite{Jureczko:Using} with the help of two existing tools (BugInfo and Ckjm). According to the authors, errors inevitably existed in the process of defect identification. Unfortunately, there may be missing links based on incomplete links between the bug database and source code repositories as illustrated in some studies \cite{Bachmann:The,Wu:ReLink}. However, these data sets collected from the PROMISE repository have been validated and applied to several prior studies. Therefore, we believe that our results are credible and suitable for other open-source projects.

Another construct threat is that, we compared the performance obtained from the predictors built with \emph{ALL}, \emph{FILTER} and \emph{TOP5} metric set for \emph{RQ2}. We recognize that this may be biased because other combinations of metrics could also achieve the best precision results. Nevertheless, the usage of these metrics, available as an oracle for comparison, is feasible because some prior studies have used these metrics to predict defect-prone classes based on the same data sets \cite{He:An}.

Threats to \emph{internal validity} concern any confounding factor that could influence our results, and they are mainly related to various measurement settings in our study. For our experiments, we choose the Top 5 metrics to build a predictor based on Equation (\ref{Eq.2}) and the results of the feature selection technique in Weka. However, we are aware that the results would change if we use a different \emph{k} length.

Threats to \emph{conclusion validity } concern the relationship between treatment and outcome, where we appropriately used a non-parametric statistical test (the Wilcoxon signed-rank) and one-way ANOVA to show statistical significance for the obtained results in \emph{RQ2} and \emph{RQ4}, respectively. Typically, one-way ANOVA is used to test for the difference among at least three groups because the two-group case can be covered by a $t$-test. In Table \ref{anova}, we analyzed the $TOP5$ metric subset and the minimum metric subset in each scenario to prove that our experiment is reliable. Additionally, it has been shown that quantitative studies aiming at statistics should test at least 20 samples to obtain statistically significant numbers; tight confidence intervals require larger samples \cite{Nielsen: How}. Fortunately,  we conducted a non-parametric test on 24 data sets and validated that the \emph{null hypothesis} in our experiments could not be rejected in most cases.

The selection of the thresholds of recall and precision is actually based on some previous studies in this field, and we have to admit that it is not a strict criterion compared with the one used in \cite{Zimmermann:Cross}. The choice of evaluation criteria depends largely on the defect data sets at hand. In general, the data sets used in our study are similar to those used in previous studies \cite{Zhang:An,He:An}. Due to the difference on software metrics between the defect data sets, the criterion used in \cite{Zimmermann:Cross} may be unsuitable for our study. On the other hand, Menzies \emph{et al.} \cite{Menzies:Defect} argue that defect predictors learned from static code metrics will reach the upper limit of performance since the information provided by static code metrics is limited. That is, it is hard for the predictors built in our context to achieve very high performance. Therefore, a simple comparison of high performance between different methods regardless of defect data sets does not make much sense.

Threats to \emph{external validity} concern the generalization of the results obtained. The main threat could be related to the selected data sets\textemdash in addition to the PROMISE repository\textemdash to validate the results of the proposed research questions. The releases are chosen from a very small subset of all projects, and there are many other public on-line data sets used for defect prediction, such as Apache and Mozilla. However, similar trends have been shown in  prior studies, which used the data sets from several code repositories. Nevertheless, we believe that our results can be reproduced using other data sets.

Another threat to \emph{external validity} concerns the choice of software metrics used to construct predictors. Although we used only static code metrics available in the literature \cite{Zimmermann:Predicting,Premraj:Network,Tosun:Validation,Nagappan:Using,Shin: Evaluating}, we are aware that other types of software metrics could exhibit different results. However, our main goal is to investigate the contribution of a simplified metric set to defect prediction from a perspective of the trade-off between generality, cost and accuracy, rather than to compare the performance of predictors built with different types of software metrics.

Last but not the least, we selected only 10 Java open-source projects, of which nine projects are developed and maintained by the Apache Software Foundation. We selected Java open-source projects because we have expertise in Java language and acknowledge the limitation of well-recognized data sets available on the Internet. The comparison of the performance of different types of software has been previously reported \cite{Zimmermann:Cross,Tosun:Validation,Menzies:Defect}.

\section{Conclusion}
This study reports an empirical study aimed at investigating how a predictor based on a simplified metric set is built and used for both WPDP and CPDP. The study has been conducted on 34 releases of 10 open-source projects available at the PROMISE repository and consists of (1) a selection of training data sets within a project and across projects, (2) a simplification of software metric set, and (3) a determination of the ability of the predictor built with the simplified metric set to provide acceptable prediction results compared with other predictors with higher complexity.

The results indicate that WPDP models capture higher precision than CPDP models, which, in turn, achieve high recall or F-measure. Specifically, the choice of training data should depend on the specific requirement of accuracy. The predictor built with a simplified metric set performed well, and there were no significant differences between our predictor and other benchmark predictors. In addition, our results also show that simple classifiers such as Na\"{\i}ve Bayes are more suitable to be the classifier for defect prediction with a simple predictor. Based on the specific requirements for complexity, generality and accuracy, the minimum metric subset is ideal because of its ability to provide good results in different scenarios and being independent of classifiers.

In summary, our results show that a simplified metric set for defect prediction is viable and practical. The prediction model constructed with a simplified or minimum subset of software metrics can provide a satisfactory performance. We believe that our metric set can be helpful for software engineers when fewer efforts are required to build a suitable predictor for their new projects. We expect some of our insightful findings to improve the development and maintenance activities.

Our future work will focus primarily on two aspects: (1) collect more open-source projects, as stated previously, to validate the generality of our approach; (2) consider the number of defects to provide an effective method for defect prediction.


\section*{Acknowledgment}
We greatly appreciate the constructive comments and useful suggestions from the editor and anonymous reviewers, which help us improve the quality and readability of our paper.

This work is supported by the National Basic Research Program of China (No. 2014CB340401), the National Natural Science Foundation of China (Nos. 61273216, 61272111, 61202048 and 61202032), the Science and Technology Innovation Program of Hubei Province (No. 2013AAA020), the Youth Chenguang Project of Science and Technology of Wuhan City in China (No. 2014070404010232), and the open foundation of Hubei Provincial Key Laboratory of Intelligent Information Processing and Real-time Industrial System (No. znss2013B017).

\end{document}